\documentclass[conference]{IEEEtran}
\usepackage{cite}
\usepackage{amsmath,amssymb,amsfonts}
\usepackage{algorithmic}
\usepackage{graphicx}
\usepackage{textcomp}
\usepackage{xcolor}
\usepackage[hyphens]{url}

\usepackage{mathptmx}
\usepackage{comment}
\usepackage{fancyhdr}
\usepackage[normalem]{ulem}
\usepackage[hyphens]{url}
\usepackage[final]{microtype}
\usepackage[keeplastbox]{flushend}
\usepackage{enumitem,kantlipsum}
\usepackage{import} 
\usepackage{mathtools}
\usepackage{comment}
\usepackage{lipsum}
\usepackage{soul}
\usepackage{graphicx}
\usepackage{multirow}
\usepackage{enumitem}
\usepackage{indentfirst}
\usepackage{authblk}

\def\BibTeX{{\rm B\kern-.05em{\sc i\kern-.025em b}\kern-.08em
    T\kern-.1667em\lower.7ex\hbox{E}\kern-.125emX}}

\newcommand{\AP}[1]{{\color{orange}{[}\textbf{AP: #1}{]}}}
\newcommand{\HA}[1]{{\color{brown}{[}\textbf{HA: #1}{]}}}

\newcommand{\JS}[1]{{\color{cyan}{[}{JS: #1}{]}}}
\newcommand{\LL}[1]{{\color{teal}{[}\textbf{LL: #1}{]}}}
\newcommand{\TODO}[1]{{\color{red}{[}{TODO: #1}{]}}}

\newcommand{\ins}[1]{{\color{black}{#1}}}

\usepackage{amsthm}
\theoremstyle{definition}
\newtheorem{definition}{Definition}[section]

\usepackage{pifont}
\newcommand{\cmark}{\ding{51}}%
\newcommand{\xmark}{\ding{55}}%

\title{Griffin: Rethinking Sparse Optimization \\ for Deep Learning Architectures} 
\author{\vspace{20pt}Jong Hoon Shin}
\author{Ali Shafiee}
\author{Ardavan Pedram}
\author{Hamzah Abdel-Aziz}
\author{Ling Li}
\author{Joseph Hassoun\vspace{-20pt}}

\affil{\em{Samsung Semiconductor Inc., San Jose, CA}}
\affil{\{jhshin.1, ali.shafiee, ardavan.p, hamzah.a, ling.li, j.hassoun\}@samsung.com\vspace{-10pt}}

\begin{document}
\maketitle
\thispagestyle{plain}
\pagestyle{plain}


\begin{abstract}

This paper examines the design space trade-offs of DNNs accelerators aiming to achieve competitive performance and efficiency metrics for all four combinations of dense or sparse activation/weight tensors.
To do so, we systematically examine the overheads of supporting sparsity on top of an optimized dense core. 
These overheads are modeled based on parameters that indicate how a multiplier can borrow a nonzero operation from the neighboring multipliers or future cycles.
As a result of this exploration, we identify a few promising designs that perform better than prior work.
Our findings suggest that even the best design targeting dual sparsity yields a 20\%-30\% drop in power efficiency when performing on single sparse models, i.e., those with only sparse weight or sparse activation tensors.
We found that one can reuse resources of the same core to maintain high performance and efficiency when running single sparsity or dense models. We call this hybrid architecture \textit{Griffin}.
Griffin is 1.2, 3.0, 3.1, and 1.4$\times$ more power-efficient than state-of-the-art sparse architectures, for dense, weight-only sparse, activation-only sparse, and dual sparse models, respectively.

\end{abstract}

\section{Introduction}
\label{sec:Intro}

In deep neural networks (DNNs)~\cite{lecun2015deep}, rectified linear unit (ReLU)~\cite{alexnet} and weight pruning~\cite{DeepCompression, han2015learning, HassibiBabak1992Pruning, Yann1989BrainDamage } enable accelerators to mitigate ineffectual computations~(i.e., those with at least one zero operand)~\cite{ineffectualsource}.
The former approach induces sparsity in activation tensors by zeroing out negative elements.
While the latter approach induces sparsity in weight tensors by pruning insignificant weights.
Although both approaches have shown promising results in several applications~\cite{niu201926ms,wu2019pruneingedge,ma2020mobile}, they might not be always enabled. 
For instance, to improve DNNs accuracy, DNN developers might prefer dense non-linear activation functions, which don't result in having as many zeros, such as swish~\cite{ramachandran2017swish}, GeLU~\cite{gelu} or leaky ReLU~\cite{maas2013rectifier}.
Similarly, they might avoid weight pruning as it significantly increases training time, drops network accuracy, or because the network is already pre-trained dense. 
Therefore, both activation and weight tensors can be dense or sparse, categorizing DNN models and execution modes in four categories based on (activation, weight) tensor types: (dense,dense), (dense, sparse), (sparse,dense), and (sparse,sparse).


An accelerator might be optimized specifically for each of the above categories.
However, DNN model categories are usually unknown at the design time for inference accelerators and might switch between different modes during training.
Unfortunately, the optimal design point to run a category of DNNs is only optimal for the same type of DNN models.
For example, architectures that are optimized for weight-only sparsity are not as efficient for activation-only sparsity models (i.e., (sparse, dense))  and cannot fully take advantage of dual sparse models (i.e., (sparse, sparse)).
Even architectures that support dual sparsity do not get the best area and power efficiency for (dense, sparse) or (sparse, dense) models (see Section~\ref{sec:Analysis}).

At the edge, where area and power have strict budgets, it is challenging to support all categories of workloads efficiently, especially when both compute unit and SRAM are optimized for a specific category so that the share of sparse overheads on top of the dense design becomes significant~\cite{Dark_Mem, SystemArdavan}.
Examples of such dense accelerators include ARM Ethos-N77 (1MB per 2048 MACs)~\cite{armEthosN77}, NVDLA (0.5MB per 1024 MACs)~\cite{nvdla}.
For instance, in the case of DNN models with (sparse, sparse) tensors, an accelerator with significant sparsity overheads to gain substantial speedup is justifiable~\cite{Sparten, Zena}. However, it does not perform efficiently for other categories.

Unlike some of the previous work~\cite{Sparten, SCNN}, we consider the possibility of efficiently running all categories of DNN execution modes, early, at the design time. 
We start with an efficient dense baseline that exploits high degrees of parallelism and data locality. 
Then we identify sources of borrowing effectual operations from future computations across several dimensions of blocking and time, and mainly exploited resources (e.g., multiplexers and buffers) that can be re-purposed from one category of execution mode to another.
We devise a novel hybrid architecture to reuse the hardware overheads in dual sparse architectures for DNN models that are only sparse in either activation or weight tensors.
We call this hybrid architecture \emph{Griffin} which enhances dual sparse architectures for both weight-only and activation-only sparse DNN models
thus maintaining power and area efficiency.
We corroborate the effectiveness of our approach by comparing our design with state-of-the-art architectures for all model categories.

To reach an optimal design for each category as well as an efficient hybrid architecture, we create an analytical model, verified by a simulator, to investigate the sources of overheads based on parameters that indicate how a multiplier can borrow a nonzero operation from the adjacent multipliers or future cycles.
Our model offers a framework to quantify several sparse architectures including some of the prior work such as Cnvlutin~\cite{Cnvlutin}, Cambricon-X~\cite{CambriconX}, and Bit-Tactical~\cite{BitTactical}. 
We further identify new design points that are more power and area-efficient than the prior state-of-the-art architectures. 
Our best designs for weight-only sparsity,  activation-only sparsity, and dual sparsity are 47\%, 223\%, and 42\% more power-efficient than prior architectures, respectively.
This model enables us to further re-purpose the logic overheads to design a hybrid architecture that is the top performer in all sparse categories. 
In summary, this paper makes the following contributions:
\begin{enumerate}[ labelwidth=!, labelindent=2pt, noitemsep]  
    \item A sparse architecture model based on how far in time and space a multiplier can borrow non-zero operands to replace with zero inputs. 
    \item A design space exploration of sparse architectures that encapsulates previous work and identifies more efficient designs for each category of networks.
    \item Techniques to reuse the logic in a dual sparse architecture to create a hybrid architecture (Griffin) so that it remains a top performer for sparse networks.
    \item We evaluate Griffin against previous optimal works. Griffin maintains a minimum of 8\% performance advantage over previous optimal works.
\end{enumerate}

The rest of this paper is organized as follows. 
Section~\ref{sec:background} defines the problem and base dense architecture.
Section~\ref{sec:Analysis} describes architectures that only support weight or activation sparsity.
Section~\ref{sec:proposed} explores supporting both activation and weight sparsity. 
Section~\ref{sec:Exp} presents our methodology and Section~\ref{sec:Results} presents the results.
Section~\ref{sec:Related} goes over the prior work.
Finally, we conclude this paper in Section~\ref{sec:conclusion}.

\begin{figure}[t!]
\vspace{-10pt}
\begin{center}
   \includegraphics[width=0.9\linewidth]{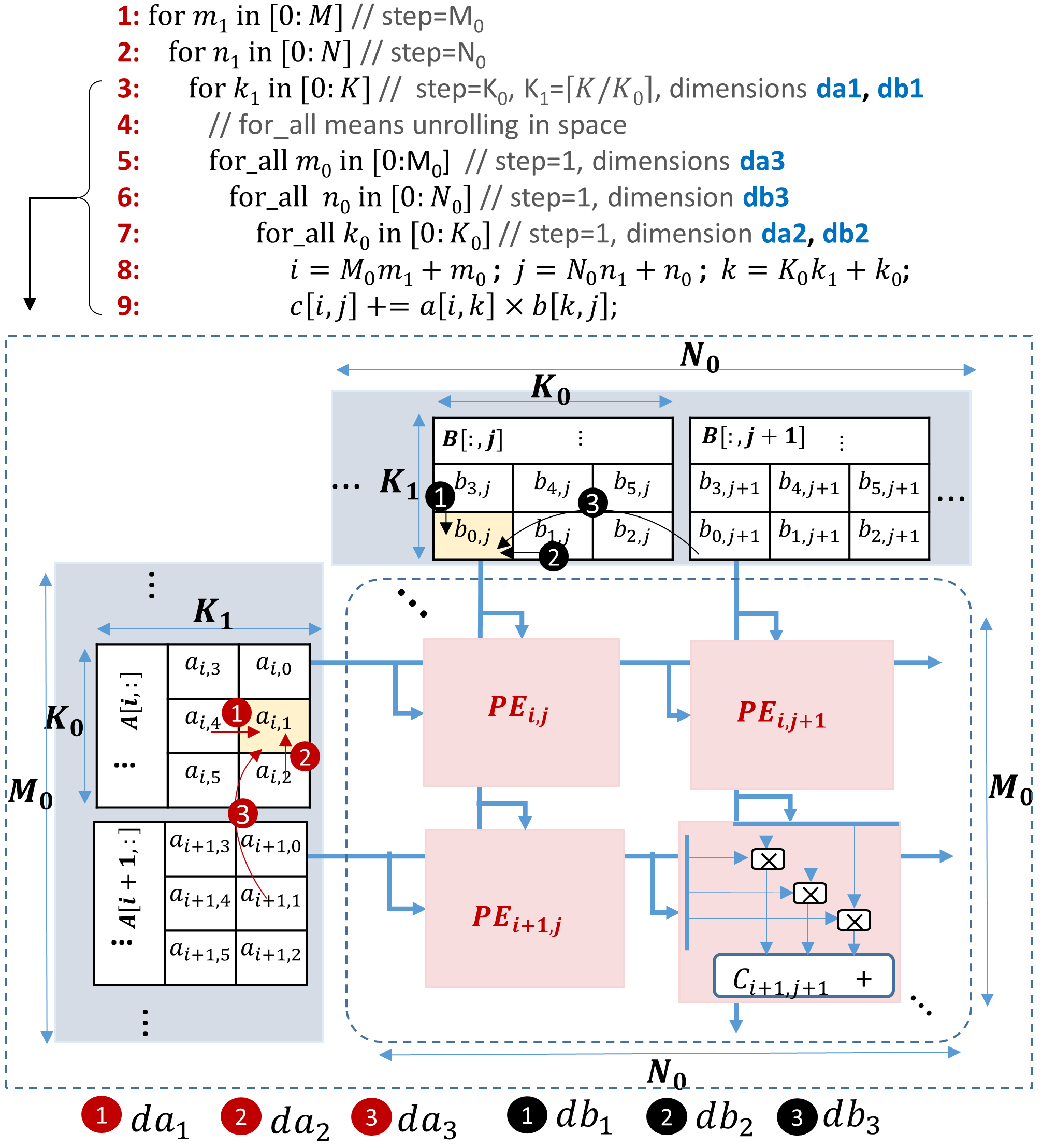}
\end{center}
\vspace{-10pt}
    \caption{
        The pseudo-code depicting the blocking of $C+=A\times B$ with $A_{M\times K}$, $B_{K\times N}$, and $C_{M\times N}$ matrices.
        The first two loops (Line 1-2) tile the codes so that it fits on the GEMM accelerator core. The accelerator realizes the inner loops (lines 3-9) and unrolls the operation in three dimension ($M_0$, $N_0$, $K_0$).  The core requires that matrices $A$ and $B$ to be reshaped as 3D tensors. The maximum distance from which a non-zero element can be borrowed across each of the dimensions of these input tensors are represented as $da_1,da_2, da_3$ for $A$ and $db_1,db_2,db_3$ for $B$.
    }
\vspace{-15pt}
\label{fig:base}
\end{figure}

\section{Problem Definition}
\label{sec:background}
In this paper, we focus on architectures that support both sparse and dense DNN models, efficiently.
We first review an optimized dense architecture in Section~\ref{sec:dense}.
Then we review different types of sparse architectures.

\subsection{Dense Architecture}
\label{sec:dense}

Most dense DNN accelerators rely on a customized unit for general matrix-matrix multiplication~(GEMM).  
GEMM, defined as $C +=A\times B$, is the main building block of popular DNNs such as  CNNs~\cite{lenet,MISHKIN201711} and Transformers~\cite{devlin2019bert}. 
For layers such as convolution layer (CL) and fully connected layer (FL), input tensor, layer's parameters (weights), and output tensor are represented as $A_{M\times K}$, $B_{K\times N}$, and $C_{M\times N}$, respectively.

In FL the kernel is represented as 2D matrix $B_{K\times N}$, and the input activations as vectors of length $K$. A batch of input activations can therefore be represented as a 2D matrix $A_{M\times K}$ with $M= Batch \ size$ that after multiplication by kernels results into a batch of outputs $C_{M\times N}$.
In a convolution layer, the kernel is represented as a 2D matrix $B_{K\times N}$ with $K = C_{in}\times R \times S$ and $N = C_{out}$, where $C_{in}$, $R$, $S$, and $C_{out}$ are the number of input channels, filter height, filter width and the number of output channels, respectively.
As a result, the input feature map is reshaped as a 2D matrix $A_{M\times K}$ with $M= H_{in} \times W_{in}$ and $ K= C_{in} \times R \times S$, where $H_{in}$ and $W_{in}$ are the height and width of each input channel.~\cite{caffe_con_troll}
For transformer-based models, GEMM operations appear in the self-attention and feed-forward layers. 
The self-attention layer leverages GEMM operations to transform token vectors to key, query, and value vectors.
Moreover, checking the similarity between all the generated query and key vectors is performed by GEMM.

GEMM can be implemented in hardware with two main optimizations:
(1) Memory hierarchy optimization using blocking to minimize the size and the data movement between different levels of the memory hierarchy ~\cite{LAP,Dark_Mem,SystemArdavan, siustuart2018memory}, and 
(2) unrolling the nested loops in space to exploit parallelism and minimize energy per access~\cite{LAP,Dark_Mem,tensorcore, kwon2020maestro, SystemArdavan, yang2020interstellar}.
Figure~\ref{fig:base} illustrates the high-level structure of the dense GEMM accelerator and how the operation is mapped onto it. 
As shown in the figure, the 3D realization of GEMM requires both matrix A and B to be rearranged/blocked in three dimensions as well.  
More specifically, each rows/columns of Matrix A/B, respectively, is stored in a 2D fashion in SRAM bank. 
Therefore, each element in A/B is adjacent to other elements in three dimensions as shown in Figure~\ref{fig:base} for $a_{i,1}$ and $b_{0,j}$. 
The adjacency of elements in the three dimensions is key to our approach of modeling sparse architecture described in Section~\ref{sec:Analysis} and Section~\ref{sec:proposed}.
More specifically an element can be borrowed from maximum distance of $da_1$, $da_2$ and $da_3$ ($db_1$, $db_2$, and $db_3$) across all neighboring dimensions.

\begin{figure*}[tp!]
\vspace{-10pt}
\begin{center}
   \includegraphics[width=0.90 \linewidth]{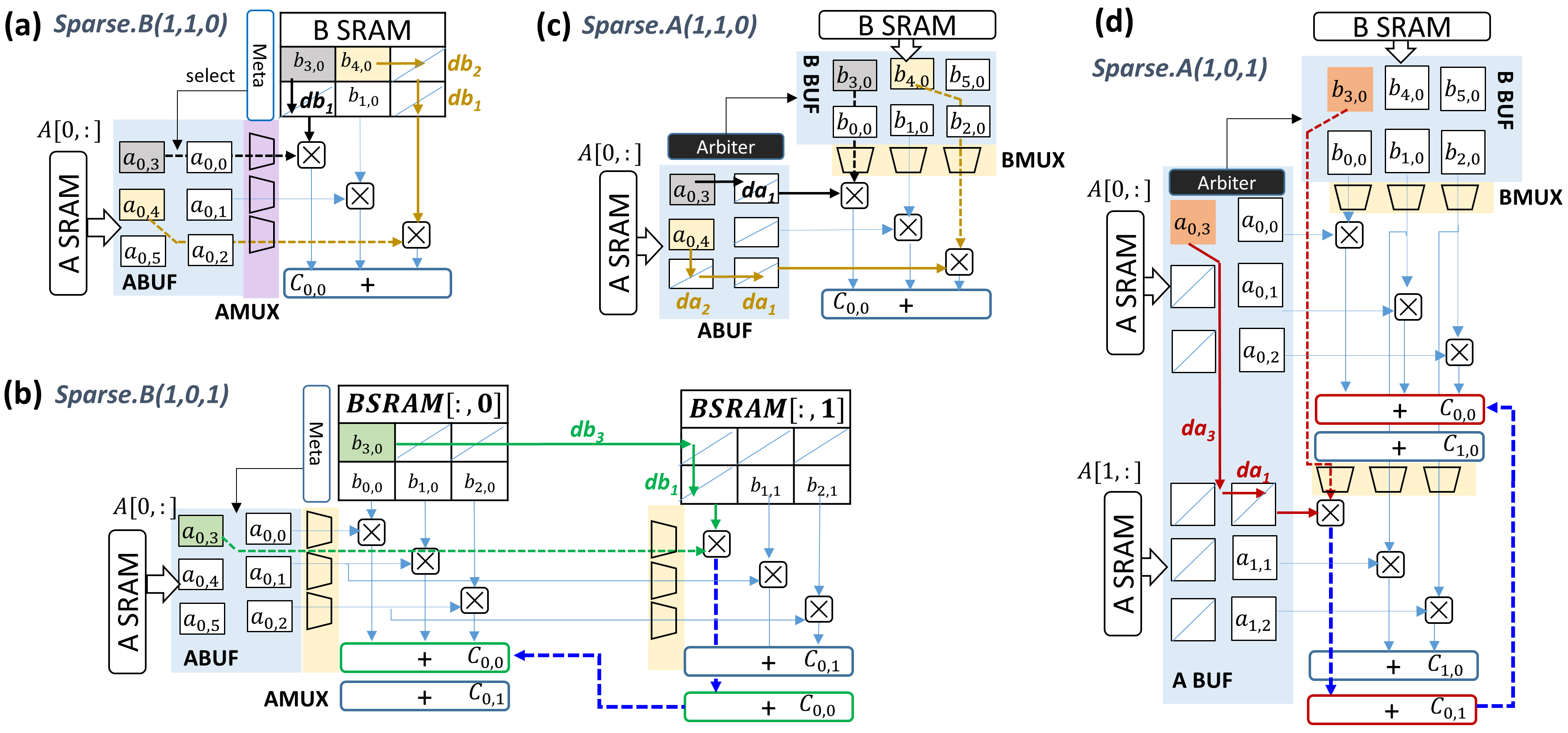}
\end{center}
\vspace{-15pt}
   \caption{
   The core replacing zero operations by borrowing from different dimensions. During preprocessing, the zero entities in B are filled with nonzero values from neighboring elements (a) in $db_1$ and $db_2$ dimensions (solid gray: borrowing from $db_1$, solid yellow: borrowing from both $db_1$ and $db_2$, \ins{dashed: selecting the associated A-elements}), (b) from $db_3$ dimension (solid green: borrowing from both $db_1$ and $db_3$, \ins{dashed green/blue: selecting and sending back the associated A-element and partial output, respectively.}). The metadata from preprocessing is later used to select appropriate values from Matrix A. The zero entities on Matrix A is detected and skipped on-the-fly. To avoid under-utilization, zero operations are replaced from neighboring entities (c) in $da_1$ and $da_2$ dimensions (gray: borrowing from $da_1$, yellow: borrowing from both $da_1$ and $da_2$, \ins{dashed: selecting the associated B-elements}) and (d) in $da_3$ dimension (orange: borrowing from both $db_1$ and $db_3$, \ins{dashed orange/blue: selecting and sending back the associated B-element and partial output, respectively.}).
}
\vspace{-15pt}
\label{fig:Weight}
\end{figure*}

\subsection{Sparse Models and Architectures}
\label{sec:definition}


There are four categories of DNN models as described in Section~\ref{sec:Intro}. \ins{Sparsity of A/B  in DNN.dense, DNN.A, DNN.B, and DNN.AB categories are dense/dense, sparse/dense, dense/sparse, sparse/sparse, respectively.}
Analogously, there are four categories of accelerators, \ins{(Dense, Sparse.A, Sparse.B, Sparse.AB)} that each is optimized for its corresponding model as summarized in Table~\ref{tab:table0}.
We also refer to \emph{hybrid~(Griffin) architecture} as the one optimized for all four categories.
For all of the above architectures, the goal is to maximize power and area efficiency with minimum overhead on the dense category. 
Throughout the next two sections, we quantify the overheads and propose some novel solutions.

\begingroup
\setlength{\tabcolsep}{3pt}
\begin{table}[t]
\vspace{-0pt}
    \centering

    \caption{Benchmarks listed with their DNNs categories and optimal accelerator architecture type}
    \vspace{-7pt}

 \scriptsize
 \begin{tabular}{|c |c| c | c|} 
  \hline
Benchmarks &  \begin{tabular}{@{}c@{}}A/B\\sparsity\end{tabular}& \begin{tabular}{@{}c@{}}DNN\\Category\end{tabular} & \begin{tabular}{@{}c@{}}Optimal\\Architecture\end{tabular} \rule{0pt}{3ex} \\ 
 \hline\hline
\begin{tabular}{@{}c@{}}CNN+Non-ReLU~\cite{lenet, He_2015_ICCV, ramachandran2017swish}\\ Trasformer+GeLU~\cite{devlin2019bert,gelu}\end{tabular}  &\begin{tabular}{@{}c@{}}dense\\/dense\end{tabular}& DNN.dense & Dense \rule{0pt}{3ex}\\
\hline
\begin{tabular}{@{}c@{}}CNN+ReLU~\cite{alexnet,resnet} \\ Trasformer+ReLU~\cite{mobilebert,attention_is_all}
\end{tabular}  & \begin{tabular}{@{}c@{}}sparse\\/dense\end{tabular} & DNN.A & Sparse.A \rule{0pt}{3ex}\\
\hline
\begin{tabular}{@{}c@{}}Pruned CNN+Non-ReLU~\cite{DeepCompression} \\ Pruned Trasformer+GeLU~\cite{movementPruneBert}\end{tabular}  &\begin{tabular}{@{}c@{}}dense\\/sparse\end{tabular}& DNN.B & Sparse.B \rule{0pt}{3ex}\\
\hline
\begin{tabular}{@{}c@{}}Pruned CNN+ReLU~\cite{DeepCompression,rigl} \\ Pruned Trasformer+ReLU~\cite{prunedReluBert}\end{tabular}  & \begin{tabular}{@{}c@{}}sparse\\/sparse\end{tabular}& DNN.AB & Sparse.AB \rule{0pt}{3ex}\\
\hline
\end{tabular}
\vspace{-20pt}

    \label{tab:table0}
\end{table}
\endgroup

\section{Sparsity Overhead Analysis}
\label{sec:Analysis}
In a typical dense GEMM accelerator, multipliers share operand fetch logic and all of them execute operations concurrently~\cite{v100,tpu,diannao,DaDianNao,eyeriss,ucnn,davinci,timeloop}.
While this approach minimizes the control overhead, dense datapath is unable to skip \emph{ineffectual operations}~\cite{ineffectualsource}. 
Sparse architectures provide additional logic to find zero operands, either by preprocessing or on the fly detection and skipping of ineffectual operations. 
The skipped operations would be replaced with nonzero operations from future cycles of the same or adjacent multipliers.
The adjacent multipliers are those that their operands are within proximity elements. 
We consider two elements as close in Matrix A/B, if they are close in  $da_1$, $da_2$, and $da_3$ or $db_1$, $db_2$, and $db_3$, respectively (See Figure~\ref{fig:base}).


In general, detecting zero operations and replacing them with nonzero ones imposes additional overheads on top of the dense core. 
Figure~\ref{fig:Weight} illustrates the overheads based on the dimension of the two adjacent operands when only one of Matrix A or B are sparse.
Matrix B is known before execution, hence it is preprocessed before being written into SRAM banks.
Preprocessing replaces zero entities with nonzero entities from the neighboring elements, which generates the metadata as well as a more compressed form of Matrix B.
Figure~\ref{fig:Weight}(a) shows the case where $b_{0,0}$ and $b_{2,0}$ are zero and replaced by $b_{3,0}$ and $b_{4,0}$, respectively.
\ins{More specifically, the solid black arrow along $db_1$ dimension represents that non-zero $b_{3,0}$ is sent to the zero $b_{0,0}$. The two solid golden arrows send $b_{4,0}$ to the diagonally neighboring element $b_{2,0}$ along $db_2$ and $db_1$ dimensions. Therefore,} for the pair $(b_{0,0},b_{3,0})$ and pair $(b_{2,0},b_{4,0})$, the borrowing distances are  $(db_1,db_2,db_3)=(1,0,0)$ and $(db_1, db_2,db_3)=(1,1,0)$, respectively.
In either of these cases, extra MUXs, called \textbf{AMUX}, before operands in $A$ are needed. These MUXs select \ins{appropriate A-elements via the dashed black/golden arrows}, based on the metadata derived from matrix B.

In Figure~\ref{fig:Weight}(b), $b_{0,1}$ is replaced with $b_{3,0}$ \ins{which is sent to diagonal direction through two consecutive green arrows along $db_3$ and $db_1$ dimensions. Thus, the borrowing} distance is $(db_1,db_2,db_3)=(1,0,1)$, which leads to performing the computation in a neighboring PE (i.e., $PE_{0,1}$). Therefore, \ins{AMUX and extra adder tree are required to navigate the associated A-element to the multiplier (dashed green arrow) and the partial result back to the accumulator in $PE_{0,0}$ (dashed blue arrows), respectively.}

In contrast to Matrix B, Matrix A is not preprocessed before execution, thus it requires performing on-the-fly zero operands detection as well as replacing them with nonzero ones.
Figure~\ref{fig:Weight}(c) depicts two cases where zero operands $a_{0,0}$ and $a_{0,2}$ are replaced with nonzero operand $a_{0,3}$ and $a_{0,4}$, respectively.
The distance for pair $(a_{0,0},a_{0,3})$ and ($a_{0,2},a_{0,4}$) are respectively $(da_1,da_2,da_3)=(1,0,0)$ \ins{(solid black arrow)} and $(da_1,da_2,da_3)=(1,1,0)$ \ins{(solid golden arrow)}.
In this case, an arbiter logic is needed to detect and replace zero operands.
At any cycle, the arbiter looks into a window of elements of $A$ that are fetched from SRAM and currently reside in a buffer called \textbf{ABUF}. 
The elements of B corresponding to elements of A in ABUF are also fetched to a buffer called \textbf{BBUF}. 
The arbiter selects nonzero operands in ABUF and generates the indices to pick appropriate values from BBUF.
These indices are fed into MUXs after BBUF, called \textbf{BMUX}\ins{, to select associated elements of B, $b_{3,0}$ and $b_{4,0}$, according to the dashed black arrow and the dashed golden arrow, respectively.}. 
Note that ABUF also requires MUXs, but they can be shared between all PEs in a row, while each PE requires a standalone BMUX.

Figure~\ref{fig:Weight}(d) illustrates a case that $a_{1,0}$ is replaced with $a_{0,3}$ \ins{by two consecutive solid orange arrows along $da_3$ and $da_1$ dimensions (Therefore,} distance is $(da_1,da_2,da_3)=(1,0,1))$. \ins{The associated B-element, $b_{3,0}$, in BBUF is selected by BMUX. (dashed orange arrow)}
In this case, an extra adder tree is required to send partial-sum values to the correct accumulator (accumulator in $PE_{0,0}$), as the multiplication is performed in the adjacent PE (i.e. $PE_{0,0}$) with a different accumulator \ins{(dashed blue arrows)}.
When only one of the input matrices is sparse, we can define single sparse architectures based on the maximum distance of borrowing nonzero operands in each dimension to replace zero operands with nonzero ones.

\begin{definition}
$Sparse.A(da_1,da_2,da_3)$ is an architecture that only supports sparsity in matrix $A$ of the GEMM operation. This architecture allows replacing a zero value $(x_1,x_2,x_3)$ with a nonzero value $(x_1+\Delta_1,x_2+\Delta_2,x_3+\Delta_3)$ where $\Delta_i\leq da_i$.
\end{definition}

\begin{definition}
$Sparse.B(db_1,db_2,db_3)$ is an architecture that only supports sparsity in matrix $B$ using preprocessing. This architecture allows replacing a zero elements $(x_1,x_2,x_3)$ with a nonzero element $(x_1+\Delta_1,x_2+\Delta_2,x_3+\Delta_3)$ where $\Delta_i\leq db_i$.
\end{definition}

\ins{Therefore, the single sparse architectures described in Figure~\ref{fig:Weight}(a), (b), (c), and (d) can be specified as Sparse.B(1,1,0), Sparse.B(1,0,1), Sparse.A(1,1,0), and Sparse.A(1,0,1), respectively.} In the architectures mentioned above, there are $da_1\times da_2\times da_3$ and $db_1 \times db_2 \times db3$ potential candidates for replacing a zero operand in A and B, respectively. In this work, we use a similar priority mechanism as~\cite{BitTactical} whenever multiple nonzero candidates exist. 
For both $Sparse.A$ and $Sparse.B$ families of architectures, we recognize the following sources of overhead: ABUF, AMUX, BBUF, BMUX, and adder tree~(ADT).
The depth of ABUF and BBUF, the fan-in of AMUX and BMUX, and the number of required adder trees depend on the limits of distance for replacement elements in different dimensions.
Table~\ref{tab:overheads} expresses the dependency for both $Sparse.A$ and $Sparse.B$.  In this table, we also provide other restricted instances from each family of architectures to reveal the impact of replacing elements in each direction. \ins{Using equations in Table~\ref{tab:overheads}, we can directly estimate and compare the cost of different single sparse architectures. For example, Sparse.A(1,1,0) shown in Figure~\ref{fig:Weight}(d) can be upgraded to Sparse.A(1,1,1) by enabling $da_3=1$, which requires twice larger AMUX fan-in,  and one extra adder-tree per PE.}

\begin{table}[t]
   
    \centering
    \vspace{-10pt}
    \caption{Hardware overhead for $Sparse.A$ and $Sparse.B$ family of architectures. The special cases show the overhead of borrowing from each direction. The unit for buffers and MUXs is the number of words.}
    \vspace{-7pt}
    
\setlength{\tabcolsep}{2pt} 
\resizebox{1\columnwidth}{!}{%
\renewcommand{\arraystretch}{1.4}
\tiny
 \begin{tabular}{|c || c| c| c | c | c|} 
 
  \hline
 Architecture & ABUF & AMUX & BBUF & BMUX & ADT  \rule{0pt}{2ex}\\ 
   
   & (depth) & (fan-in) & (depth) & (fan-in) & (number)
 \rule{0pt}{2ex}\\ 
 \hline\hline
 \multicolumn{6}{|c|}{Operand selection via metadata} \rule{0pt}{2ex}\\
 \hline
 $Sparse.A(da_1,0,0)$ & $1+da_1$ & $1+da_1$ & $1+da_1$ & $1+da_1$  & 1 \rule{0pt}{2ex}\\ 
 \hline
 $Sparse.A(1,da_2,0)$ & 2 & $2+da_2$ & 2 & $2+da_2$  & 1  \rule{0pt}{2ex}\\ 
 \hline
 $Sparse.A(1,0,da_3)$ & 2 & $2+da_3$ & 2 & 2  & $1+da_3$ \rule{0pt}{2ex}\\ 
 \hline
 $Sparse.A$ & $1+da_1$ & $1+da_1$ & $1+da_1$ & $1+da_1$ & $1+da_3$  \rule{0pt}{2ex}\\
 $(da_1,da_2,da_3)$ & & $\times(1+da_2)$ & & $\times(1+da_2)$ & \rule{0pt}{2ex}\\
  & & $\times(1+da_3)$ & &  & \rule{0pt}{2ex}\\
 \hline
 $Sparse.B(db_1,0,0)$ & $1+db_1$ & $1+db_1$ & - & -  & 1 \rule{0pt}{2ex}\\ 
 \hline
 $Sparse.B(1,db_2,0)$ & 2 & $2+db_2$ & - & -  & 1  \rule{0pt}{2ex}\\ 
 \hline
 $Sparse.B(1,0,db_3)$ & 2 & $2$ & - & -  & $1+db_3$  \rule{0pt}{2ex}\\ 
 \hline 
 $Sparse.B$ & $1+db_1$ & $1+db_1$ & - & -  & $1+db_3$  \rule{0pt}{2ex}\\ 
 $(db_1,db_2,db_3)$ & & $\times(1+db_2)$ & & & \rule{0pt}{2ex}\\
 \hline
\end{tabular}
}
\vspace{-15pt}

    \label{tab:overheads}
\end{table}

\textbf{Load Balancing}: For unstructured sparse input matrix A and B the zero operands are not necessarily evenly distributed. This issue still exists after preprocessing $B$ or on-the-fly zero skipping on $A$.
Coarse-grain load balancing is an effective approach to distribute nonzero values and improve performance utilization~\cite{Sparten, Zena}. In this approach a GEMM operation is decomposed into smaller blocks and each block is assigned to available idle PEs.
In this work, however, we consider a recently proposed fine-grain approach by shuffling input matrix $A$ and $B$ along  their second dimension (i.e., $da_2$ and $db_2$) in the core~\cite{crane}.
The shuffling happens, over the dense matrices $A$ and $B$, before applying preprocessing or entering the buffer for on-the-fly zero skipping.
While there are many ways to perform shuffling, we observe that simple permutation is sufficient.
Thus, if an element is located in $(i_1,i_2,i_3)$ in an input matrix, it will be relocated to $(i_1, (i_2 mod K_0),i_3)$, where $K_0$ is the size of the second dot product unit ( see Figure~\ref{fig:base}).
Note that shuffling happens on both Matrices $A$ and $B$.
To navigate the elements of $A$ to its corresponding $B$'s elements, rotation-based shuffling requires a $K_0\times K_0$ crossbar between SRAM and ABUFs.
Therefore we limit the shuffling to local rotations only between four consecutive elements (in $da_2$ and $db_2$) to reduce the $K_0\times K_0$ crossbar to multiple ($K_0 / 4$) $4\times4$ crossbars.
In our experiments, this localization does not impact the load balancing.
We use the notation $shuffle=on$ and $shuffle=off$ to indicate whether an architecture support rotation-based shuffling or not.

\section{Hybrid Sparse Architecture}
\label{sec:proposed}

The architecture solution mentioned in Section ~\ref{sec:Analysis} can improve the performance and power efficiency of single sparse models (i.e., DNN.A and DNN.B models in Table~\ref{tab:table0}). 
In Section~\ref{sec:dual}, we first propose an architecture family that supports dual sparsity (i.e., DNN.AB models in Table~\ref{tab:table0}). This is achieved by enabling zero operand replacement from all of the six dimensions mentioned in Section~\ref{sec:Analysis} ($da_1$, $da_2$, $da_3$, $db_1$, $db_2$, and $db_3$).
On top of this parametric architecture, we propose a hybrid solution in Section~\ref{sec:enhance}, with an optimal design called Griffin. Compared to a dual sparse architecture, Griffin can morph into high-performance $Sparse.A$ and $Sparse.B$ architecture and therefore perform better for DNN.A and DNN.B models, respectively.

\subsection{Supporting Dual Sparsity:}
\label{sec:dual}

The main overhead when supporting dual sparsity is to fetch operations with nonzero operands in both $A$ and $B$.
Such architectures take advantage of all six dimensions for replacing zero elements. 
We define dual sparse architectures as follows:

\begin{figure}[t]
\vspace{-10pt}
\begin{center}
   \includegraphics[width=0.95\linewidth]{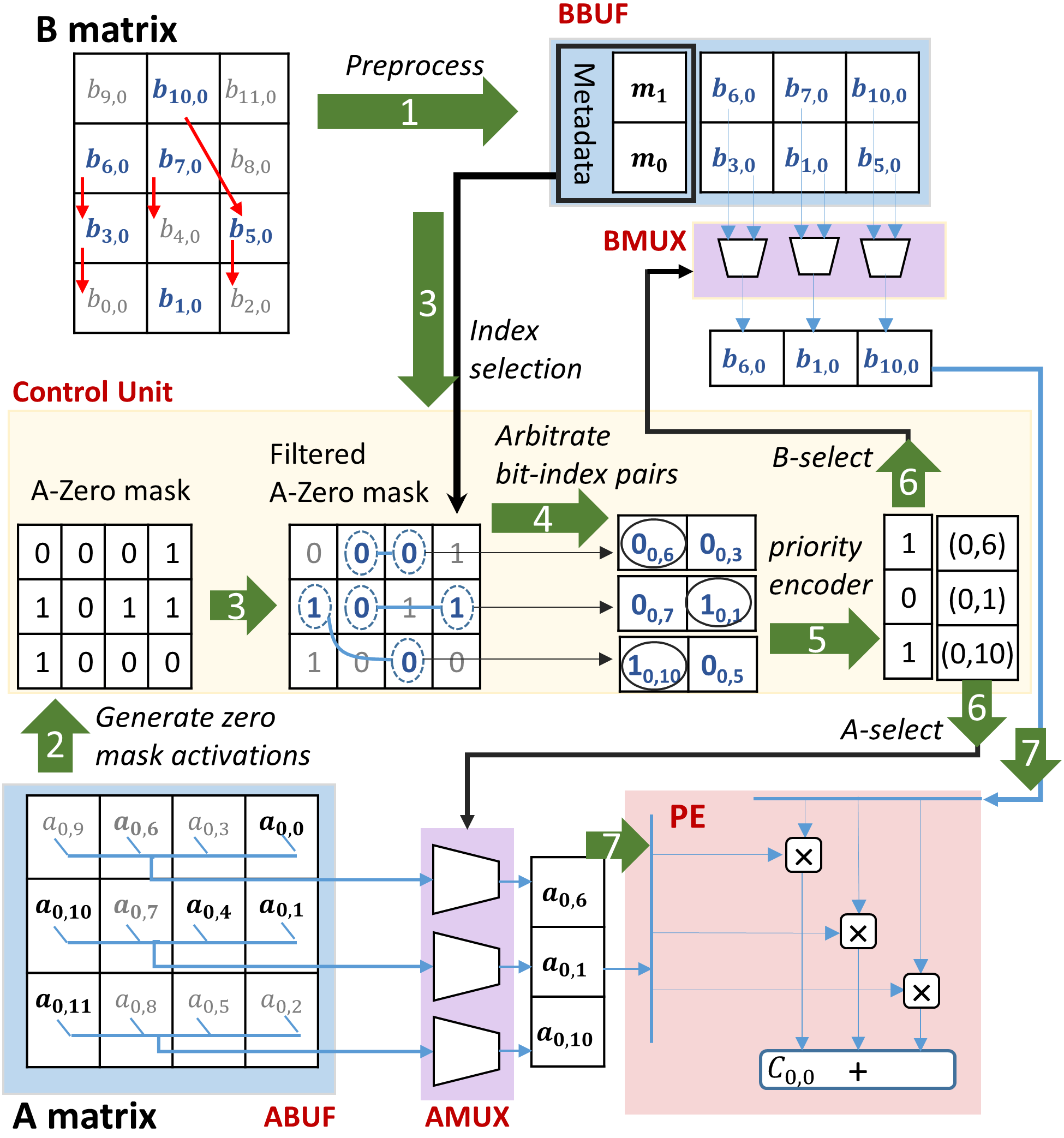}
\end{center}
\vspace{-15pt}
   \caption{
    Dual sparse architecture steps. 
    (1) Preprocessing Matrix B,
    (2) Generating zero mask for Matrix A,
    (3) Filtering out the zero mask of A,
    (4) Selecting nonzero elements,
    (5) Generating index,
    (6) Selecting elements of A and B using indices, and
    (7) Execute the dot product.
}
\vspace{-15pt}
\label{fig:Dual}
\end{figure}

\begin{definition}
$Sparse.AB(da_1,da_2,da_3,db_1,db_2,db_3)$ is an architecture that support sparsity in both  $A$ and $B$. This architecture replaces a zero operand in $A$ $(x_1,x_2,x_3)$  with nonzero value $(x_1+\Delta_1,x_2+\Delta_2,x_3+\Delta_3)$ where $\Delta_i\leq da_i$. Similarly, in Matrix B, it allows replacing a zero operand $(y_1,y_2,y_3)$  with nonzero value $(y_1+\Delta'_1,y_2+\Delta'_2,y_3+\Delta'_3)$ where $\Delta'_i\leq db_i$.
\end{definition}

In this section, we explain our approach using a walk-through example, shown in Figure~\ref{fig:Dual}.
The following seven steps are required to support dual sparsity.

\begin{enumerate}
    \item \textbf{Preprocessing $B$}: Since $B$ is known before the execution, it is preprocessed to a compressed format with metadata.
    The preprocessed elements of $B$ in SRAM are fetched in BBUF, which holds a window of current elements every cycle. In our example, $b_{1,0}$, $b_{3,0}$, $b_{5,0}$, $b_{6,0}$, $b_{7,0}$, and $b_{10,0}$ are kept in BBUF.
    \item \textbf{Generating zero masks}: ABUF keeps the elements of $A$ corresponding to the element of B currently residing in BBUF. For each of these elements in ABUF, a single mask would be generated indicating whether it is zero or not. $a_{0,0}$, $a_{0,1}$,  $a_{0,4}$, $a_{0,10}$, and $a_{0,11}$ are the elements of $A$ in ABUF with mask bit equals 1.  
    \item \textbf{Filtering zero masks}: using the metadata from $B$, the zero mask get updated by zeroing those mask bits when their corresponding weight is zero. In our example, $a_{0,0}$ , $a_{0,4}$, and $a_{0,11}$ become zero.
    \item \textbf{Arbitration}: The remaining ones in the zero mask associate with operations with both nonzero operands in $A$ and $B$. In this step, these ones are detected and selected.
    \item \textbf{Index generation}: For the selected ones in the zero mask, their indices in $A$ and $B$ are extracted using priority encoders \ins{which detect the first unused non-zero value in each bit-index pair list and generate the corresponding indices.} In our example, $a_{0,6}$ ($b_{6,0}$), $a_{0,1}$ ($b_{1,0}$), and $a_{0,10}$ ($b_{10,0}$) are selected. If there is no such pair, default value zero would be picked.
    \item \textbf{Operand selection}: Using the indices generated in Step 6, two vectors of operands are selected from ABUF and BBUF.
    \item \textbf{Execution}: The selected vectors are fed into the PE for execution.
\end{enumerate}

For a $Sparse.AB(x,y,z,x',y',z')$ the following overhead is required to realize the above 7-step processes. 
First, the control logic is used to detect nonzero operands per PE, as the pairs of A and B in each PE are different. 
Second, the ABUF is shared within a row of PEs. The depth of this buffer is $L =(1+x)\times(1+x')$. Similarly, the BBUF is shared within a column of PEs and its depth is $(1+x')$.
Third, each PE requires its own AMUX and BMUX (see Figure~\ref{fig:Dual}). The fan-in for AMUX and BMUX are $1+(L-1)\times(1+y+y')\times(1+z)$ and $1+x\times(1+y)$, respectively.
Finally, dual sparsity support requires $z\times z'$ extra adders per PE.

\subsection{Hybrid Architecture (Griffin):}
\label{sec:enhance}

A hybrid architecture reuses dual sparsity overheads to achieve better results in single sparse cases (i.e., DNN.A and DNN.B). 
In this section, we introduce, Griffin,  an optimal design of hybrid architectures with high area and power efficiency in both single and dual sparse categories.

\begin{figure}[!t]
\vspace{-10pt}
\begin{center}
   \includegraphics[width=0.93\linewidth]{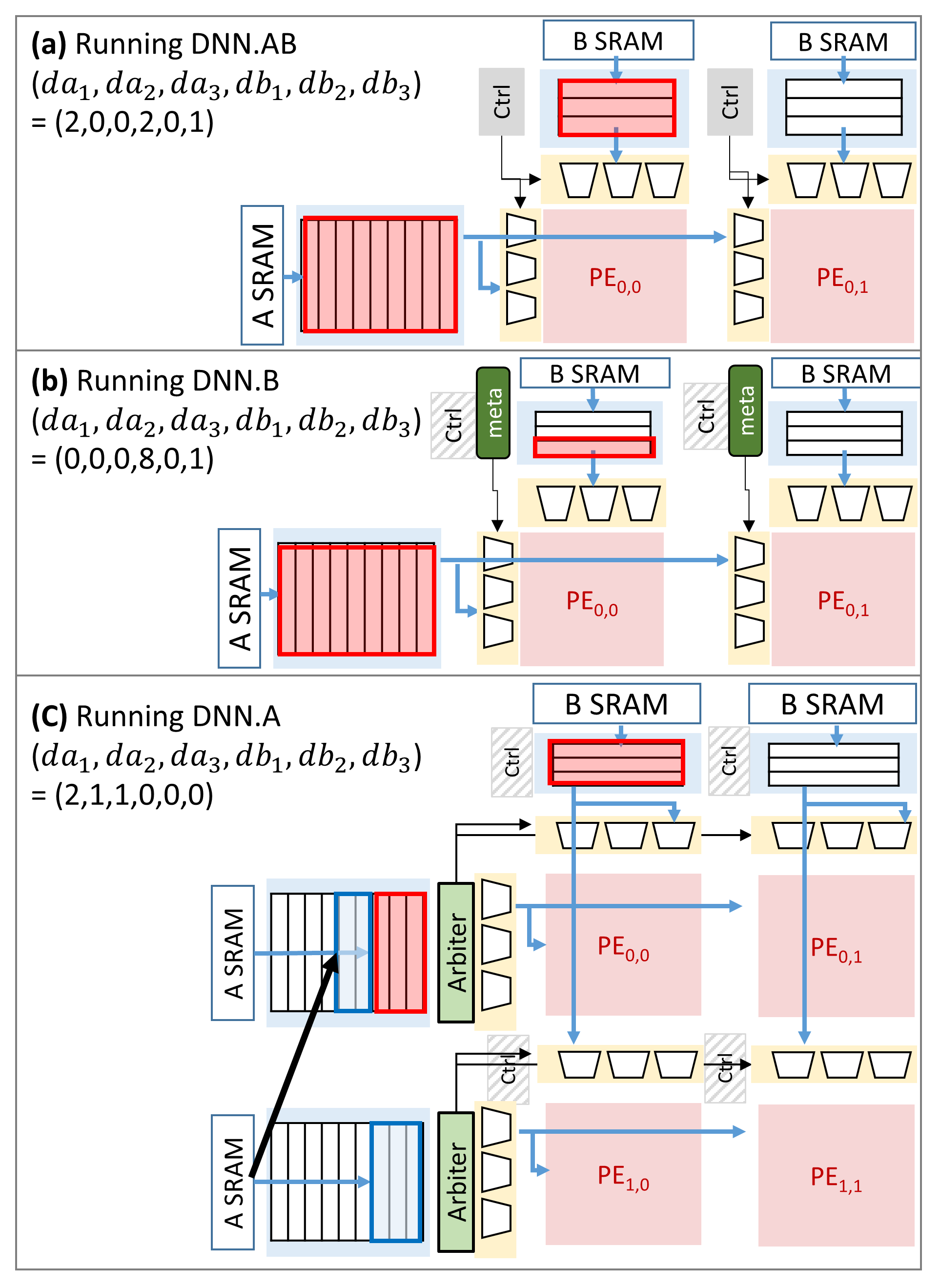}
\end{center}
\vspace{-15pt}
   \caption{
   Griffin is a hybrid architecture that morphs into different configuration for different categories of benchmarks. (a) When running dual sparse benchmark, Griffin allows (2,0,0,2,0,1) borrowing from ($da_1$, $da_2$, $da_3$, $db_1$, $db_2$, $db_3$). (b) When running benchmark with only sparse Matrix B, it morphs into Sparse.B(8,0,1). (c) When running benchmark with only sparse Matrix A, it morphs into Sparse.A(2,1,1). In each subfigures, we highlight the part of ABUF and BBUF used in the configuration for $PE_{0,0}$.
}
\vspace{-10pt}
\label{fig:enhanced}
\end{figure}

Griffin has three configurations for sparse architectures shown in Figure~\ref{fig:enhanced}. When running dual sparse benchmark, it performs as $Sparse.AB(2,0,0,2,0,1)$ (see Section~\ref{sec:dual}). 
We discuss how we derive these parameters for the optimal performances in Section~\ref{sec:Results}.
This configuration requires 9-entry ABUF, 3-entry BBUF, 9-input AMUX, and 3-input BMUXs, and one extra adder tree.
Without hybrid architecture, this design point downgrades to $Sparse.A(2,0,0)$ and $Sparse.B(2,0,1)$ for DNN.A and DNN.B models, respectively. 
In this case, the main overhead, such as the large ABUF would not be underutilized (the above-downgraded models only require 3 entries of the ABUF).
On the other hand, Griffin morphs into a more aggressive configuration for DNN.A and DNN.B models to better utilize the overhead already imposed for DNN.AB, with a negligible hardware cost. 
Specifically, Griffin uses $Sparse.A(2,1,1)$ and $Sparse.B(8,0,1)$ for DNN.A and DNN.B models, respectively.

Figure~\ref{fig:enhanced}(b) shows the re-configuration to $Sparse.B(8,0,1)$. 
This configuration uses the entire nine elements of ABUF but requires 4bits of metadata per element of $B$ rather than 3bits. 
Since Matrix $A$ is dense, the control logic in each PE is idle and the metadata values are enough to generate indices for AMUXs. 
In addition, only one entry of BBUF is used, hence BMUX indices are fixed to 0.

\begin{table}[t]
   \tiny
    \centering
    \vspace{-10pt}
    \caption{Comparison between Griffin and dual sparse $Sparse.AB(2,0,0,2,0,1)$ when running DNN.A and DNN.B.}
    \vspace{-7pt}
\resizebox{0.85\columnwidth}{!}{%
 \begin{tabular}{|c || c | c| c|| c | c | c|} 
  \hline
 & \multicolumn{3}{|c|}{Dual Sparse } & \multicolumn{3}{|c|}{Griffin }   \rule{0pt}{2ex}\\ 
 \hline\hline
 model &\multicolumn{6}{|c|}{DNN.A }  \rule{0pt}{2ex}\\
 \hline
 & \multicolumn{3}{|c|}{downgrade to } & \multicolumn{3}{|c|}{morph to } \rule{0pt}{2ex}\\
 config & \multicolumn{3}{|c|}{Sparse.A(2,0,0) } & \multicolumn{3}{|c|}{Sparse.A(2,1,1) } \rule{0pt}{2ex}\\ 
 \hline
  & \multicolumn{3}{|c|}{fan-in BMUX:3} & \multicolumn{3}{|c|}{fan-in BMUX:5} \rule{0pt}{2ex}\\
 changes/ & \multicolumn{3}{|c|}{ABUF entries used: 9 } & \multicolumn{3}{|c|}{ABUF entries used:  5} \rule{0pt}{2ex}\\
 overhead & \multicolumn{3}{|c|}{no global arbiter } & \multicolumn{3}{|c|}{one global arbiter per row} \rule{0pt}{2ex}\\
 & \multicolumn{3}{|c|}{PE controller used} & \multicolumn{3}{|c|}{PE controller not used } \rule{0pt}{2ex}\\
 \hline
 \hline
 model &\multicolumn{6}{|c|}{DNN.B }  \rule{0pt}{2ex}\\
 \hline
 & \multicolumn{3}{|c|}{downgrade to } & \multicolumn{3}{|c|}{morph to } \rule{0pt}{2ex}\\
 config & \multicolumn{3}{|c|}{Sparse.B(2,0,1) } & \multicolumn{3}{|c|}{Sparse.B(8,0,1) } \rule{0pt}{2ex}\\ 
 \hline
 changes/ &\multicolumn{3}{|c|}{BBUF entries used:3 } & \multicolumn{3}{|c|}{BBUF entries used:1 } \\
 overhead &\multicolumn{3}{|c|}{metadata per element:3b } & \multicolumn{3}{|c|}{metadata per element:4b } \rule{0pt}{2ex}\\ 
 \hline

\end{tabular}

}
\vspace{0pt}

    \label{tab:griffin_overheads}
\end{table}

Figure~\ref{fig:enhanced}(c) illustrates Griffin morphing into $Sparse.A(2,1,1)$ for DNN.A models.
In this mode, the entire three elements of BBUF are needed. 
In addition, the extra adder tree of each PE is also reused because these configurations allow borrowing from $da_3$.
However, there are three main changes to the dual sparse mode.
(1) $Sparse.A(2,1,1)$ requires three entries of ABUF from the current row and two from the neighboring one. Since ABUF has nine spaces from nine entries, the element from the neighboring ABUF is also copied into the current ABUF.
(2) The process of zero skipping and arbitration becomes more complicated as borrowing from the $da_2$ direction is permitted.
However, just one arbiter is needed per row of PE since only $A$ is sparse. Moreover, the control logic in each PE is bypassed.
(3) Due to enabling borrowing from $da_2$ direction, the fin-in of BMUXs should increase from three to five. Table~\ref{tab:griffin_overheads} summarizes the difference between Griffin and dual sparse architecture for DNN.A and DNN.B models.

\section{Experimental Setup}
\label{sec:Exp}

\begin{table}[]
\vspace{-10pt}
\centering
\caption{Summary of benchmarks and architecture configuration used for the experiments.}
\label{tab:expSetup}
\vspace{-5pt}
\setlength{\tabcolsep}{2pt} 
\resizebox{0.85\columnwidth}{!}{
\renewcommand{\arraystretch}{1.25}

\resizebox{0.80\columnwidth}{!}{%
\tiny

\begin{tabular}{|c|c|c|c|}
\hline
\multicolumn{4}{|c|}{Benchmarks}                                                                                                                                                          \\ \hline\hline
Network        & \multicolumn{1}{c|}{\begin{tabular}[c]{@{}c@{}}Sparsity ratio\\ (B,A)\end{tabular}} & Accuracy   & \begin{tabular}[c]{@{}c@{}}Dense latency\\ (\# cycles)\end{tabular} \\ \hline
AlexNet~\cite{DeepCompression}        & (89\%,53\%)                                                                         & 57.3\%~(top-1)         & $1.0\times 10^{6}$
                                                                 \\ \hline
GoogleNet~\cite{park2016faster}      & (82\%,37\%)                                                                         & 68.2\%~(top-1)           & $2.2\times 10^{6}$
                                                                 \\ \hline
ResNet50~\cite{gale2019prune}       & (81\%,43\%)                                                                         & 76.1\%~(top-1)          & $4.8\times 10^{6}$
                                                                 \\ \hline
InceptionV3~\cite{zhu2017prune}    & (79\%,46\%)                                                                         & 75.1\%~(top-1)          & $6.9\times 10^{6}$
                                                                 \\ \hline
MobileNetV2~\cite{rigl}    & (81\%,52\%)                                                                         & 67.5\%~(top-1)          & $2.2\times 10^{6}$
                                                                 \\ \hline
 BERT~(MNLI)~\cite{movementPruneBert}  & (82\%,0\%)                                                                          &   
\begin{tabular}[c]{@{}c@{}}81.0\%(Devacc)\\ 81.4\%(MMacc)\end{tabular}     &  \begin{tabular}[c]{@{}c@{}}$5.3\times 10^{6}$ \\ (sentence length=64)\end{tabular}                                                      \\ \hline\hline
\multicolumn{4}{|c|}{Architecture   Configuration}                                                                                                                                        \\ \hline\hline
Core           & \multicolumn{1}{c|}{($K_0$,$N_0$,$M_0$) =}                                                   & \#MACs/core     & 1024                                                                   \\ 
Dimension      & \multicolumn{1}{c|}{(16,16,4)}                &   &                                       \\ \hline
\# Cores       & 1                   &                                Technology     & \multicolumn{1}{c|}{7nm} \\ \hline
ASRAM        &          512kB                                                   & BSRAM        & 32kB    \\ \hline
ASRAM-BW       &  51GB/s                                                                & BSRAM-BW        & 205GB/s \\ \hline
DRAM-BW        &          50GB  &  Supply Voltage & \multicolumn{1}{c|}{0.71v}                                 \\ \hline
Target Freq. & 800MHz                                                           & Dataflow & OS                                                            \\ \hline
\end{tabular}
}
}
\vspace{-10pt}
\end{table}

The performance evaluation is based on a cycle-accurate simulation model developed in Python and PyTorch for several pruned DNN benchmarks. 
\ins{The python-based simulator receives weight and activation tensor blocks from pytorch and pre-preprocess weight tensors, if necessary, and then computes the number of cycles per block of tensors according to the borrowing strategy. Our simulation pipelines consider stalls due to output synchronization, SRAM bank conflicts, and ABUF/BBUF fullness.  
}
The cycle-accurate simulation estimate the inference end-to-end latency for the given benchmarks. 
The weight and activation (B,A) sparsity ratios, accuracy, and latency (i.e., number of cycles) with dense matrices  for these benchmarks are listed in Table~\ref{tab:expSetup}.

We consider the dense architecture depicted in Figure~\ref{fig:base} as our baseline.
The datapath configuration for this architecture is $(K_0,N_0,M_0)=(16,16,4)$ with 1024 MAC operations per cycle.
Our default multiplier precision for the MAC units is INT8. 

The baseline on-chip memory is mainly $A$ SRAM (ASRAM) and $B$ SRAM (BSRAM). 
We optimized the baseline memory for better area efficiency by allocating only 512KB for ASRAM and 32KB for BSRAM. These SRAM sizes are within the range of efficient memory hierarchy design and consistent with commercial DNN accelerators such as NVDLA (512KB per 1024 MACs)~\cite{nvdla} and ARM Ethos-N77 (1MB per 2048 MACs)~\cite{armEthosN77}. 
Note that, the area overheads to support sparsity are mostly in the computation cores. 
Thus, a large ASRAM can be misleading as its area/energy consumption overshadows the impact of sparse overheads on the cores~\cite{Dark_Mem}. For baseline architecture, ASRAM and WSRAM bandwidth are 51.2GB/s and 204.8GB/s, respectively.
To exploit the full sparsity speedup, SRAM BW should be equal or more than the multiplication of the normalized speedup and the baseline bandwidth. 
We used 50GB/s DRAM bandwidth which is enough to avoid any performance drop.

On top of the dense baselines, we also evaluate our architecture against three state-of-the-art sparse (SOTA) architectures: BitTactical~\cite{BitTactical}, TensorDash~\cite{TensorDash}, and SparTen~\cite{Sparten}. 
We denote the dual sparse architectures TensorDash and Sparten as \textbf{TDash.AB} and \textbf{SparTen.AB}, respectively.
We refer to the weight-only sparse architectures BitTactical to \textbf{TCL.B} \ins{, which can be considered as single sparsity version of TDash.AB with weight preprocessing.} We also use \textbf{SparTen.B} and \textbf{SparTen.A} to refer to one-sided SparTen optimized for DNN.B and DNN.A, respectively.
Table~\ref{tab:my-table} summarizes the SOTA architectures that we compare against. 
For a fair comparison, we implemented these SOTA architectures with the same configuration as our baseline models (e.g., 1024 8-bit MACs).
These SOTA architectures show superior performance to other prior work \cite{SCNN, CambriconX, Cnvlutin}. 

To estimate power and area, we implemented the baseline, our proposed architecture, \ins{and the three SOTA sparse architectures with the same SRAM capacity} in SystemVerilog and synthesized them using Synopsys DesignCompiler~\cite{DC} and Synopsys memory compiler with $7nm$ technology libraries. 
We consider $800$~$MHz$ clock frequency and $0.71V$ voltage for our synthesis processes. 
The experimental setup and configurations are summarized in Table~\ref{tab:expSetup}.

\begin{table}[]
\vspace{-5pt}
\caption{Comparison of routing dimensions in matrices A and B for various architectures.}
\label{tab:my-table}
\vspace{-5pt}
\setlength{\tabcolsep}{2pt} 
\resizebox{\columnwidth}{!}{%
\renewcommand{\arraystretch}{1.25}
\begin{tabular}{|l|c|c|c|c|c|c|c|c|}
\hline

\multirow{2}{*}{{\textbf{Architecture}}} & \multicolumn{3}{c|}{\textbf{A-matrix Routing}} & \multicolumn{3}{c|}{\textbf{B-matrix Routing}} & \multirow{2}{*}{\textbf{Shuffle}} & \textbf{Sparsity} \\ \cline{2-7}
                              & \textbf{$da1$}         & \textbf{$da2$}        & \textbf{$da3$}        & \textbf{$db1$}         & \textbf{$db2$}        & \textbf{$db3$}        &                          & \textbf{Support}  \\ \hline\hline

Baseline       & \xmark & \xmark & \xmark & \xmark & \xmark & \xmark & \xmark & Dense            \\ \hline
BitTactical    & \xmark & \xmark & \xmark & \cmark & \cmark & \xmark & \xmark & Weight Only      \\ \hline
TensorDash     & \cmark & \cmark & \xmark & \cmark & \cmark & \xmark & \xmark & Dual Sparsity   \\ \hline
SparTen        & \xmark & \cmark & \xmark & \xmark & \cmark & \xmark & \xmark & Dual  Sparsity  \\ \hline
Griffin (ours) & \cmark & \cmark & \cmark & \cmark & \cmark & \cmark & \cmark & Hybrid  Sparsity  \\ \hline
\end{tabular}
}
\vspace{-15pt}
\end{table}


We use the geometric mean to estimate the evaluation metrics which include normalized speedup and power/area efficiency.
For power and area efficiency metrics, we consider effective $TOPS/W$ and $TOPS/mm^2$ defined as follows.
\begin{definition}
Power and area efficiency:

Effective $TOPs/W$ = sparsity speedup$\times$dense $Tops/W$ 

Effective $TOPs/mm^2$ = sparsity speedup$\times$dense $Tops/mm^2$ 
\end{definition}




\section{Results}
\label{sec:Results}

In this section, the design space exploration results of sparse architectures such as weight-only sparse, activation-only sparse, and dual sparse architectures are reported. We identify the optimal design points for each sparse architecture class, and also propose a hybrid architecture \emph{Griffin} which performs well in three different DNN categories, DNN.A, DNN.B, and DNN.AB.

\begin{figure*}[ht!]
\vspace{-10pt}
\begin{center}
   \includegraphics[width=15.5cm]{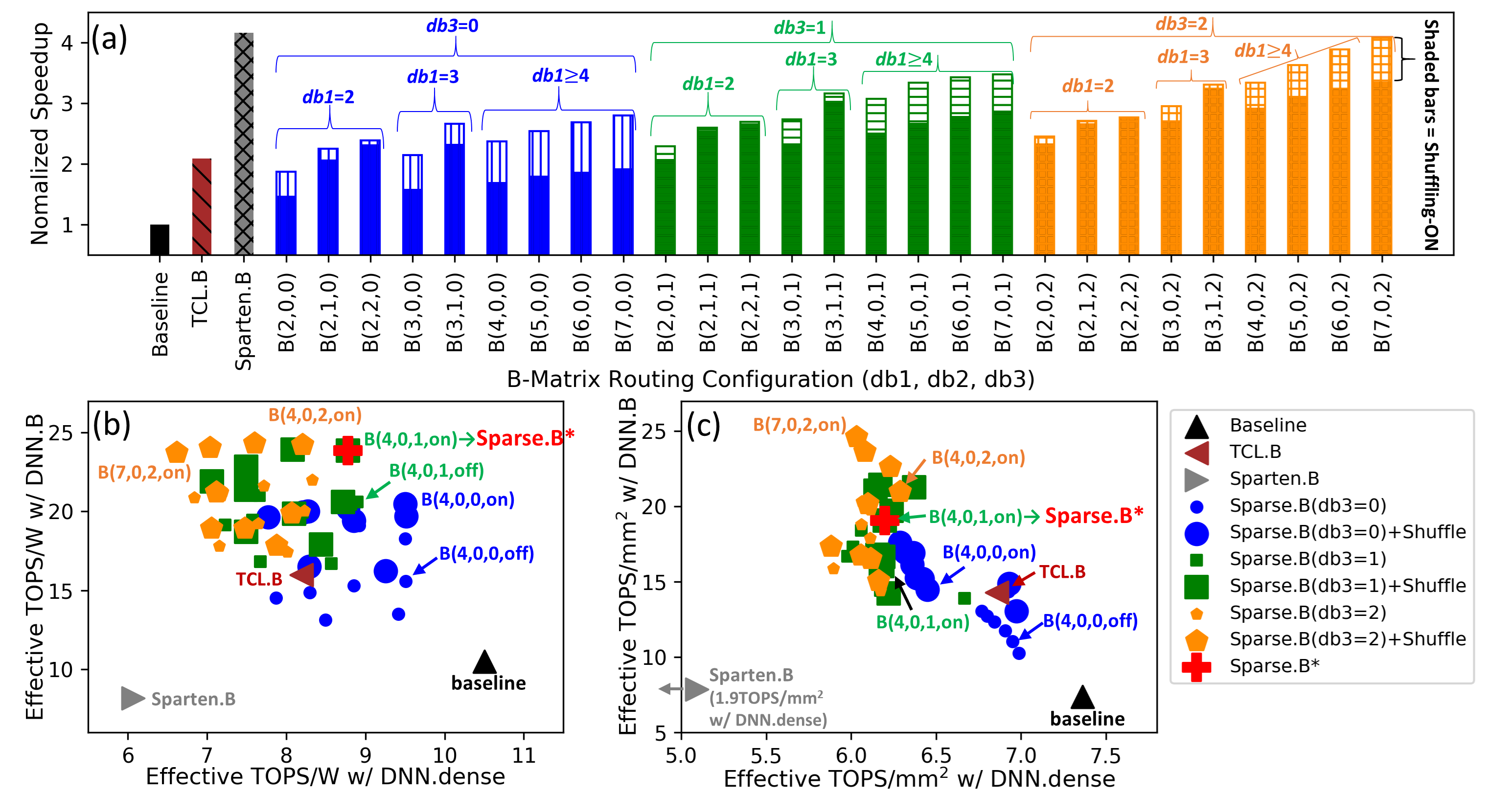}
\end{center}
\vspace{-15pt}
   \caption{
   Impact of B-matrix routing configurations. We use shorter notation B($db1$,$db2$,$db3$) to refer to Sparse.B($db1$,$db2$,$db3$). When we indicate shuffling-on/off of the sparse design, `on' or `off' index is added as Sparse.B($db1$,$db2$,$db3$,$on/off$). (a) Normalized speedup with respect to the dense baseline for different configurations. The shaded bars are the results with shuffling. 
   (b) Effective power efficiency for different design points for DNN.B benchmark (y-axis) and DNN.dense benchmark (x-axis), 
   (c) Effective area efficiency for different design points for DNN.B benchmark (y-axis) and DNN.dense benchmark (x-axis). 
   }
\vspace{-15pt}
\label{fig:W_configs}
\end{figure*}

\subsection{Weight-Only Sparsity Support:}

\indent For weight-only sparse architectures, we limit our analysis to those with AMUX fan-in smaller than or equal to eight, as larger MUXs would severely impact power efficiency. 
Figure~\ref{fig:W_configs}(a) shows normalized speed up, with respect to the dense baseline, for all possible configurations with the above constraints.
We remove the case with $db1=1$, as it is far from the optimal points. Below are the key observations from Figure~\ref{fig:W_configs}.

\begin{enumerate}[leftmargin=*, label={(\arabic*)}]
   \item Larger $db1$, $db2$, $db3$ leads to higher speed-up, with $db1$ has more impact than the other two parameters because it decides the ideal maximum speed-up to $(1+db1)$. 
  \item $db3>0$ can boost the performance by up to 48\% \\ ($db3=1$, B(4,0,0,off):1.7$\times$~$\rightarrow$~B(4,0,1,off):2.5$\times$) and\\72\% ($db3=2$, B(4,0,0,off):1.7$\times$~$\rightarrow$~B(4,0,2,off):2.9$\times$) \ins{with increasing power overehead of 10\% and 20\%, respectivley.}

  \item Shuffling is effective, mostly for $db1 >2$  and leads to up to 43\% improvement in speedup (B(6,0,0,off):1.9$\times$ $\rightarrow$ B(6,0,0,on):2.7$\times$) (See Figure~\ref{fig:W_configs}(a)).
  \item Shuffling plays similar role as $db2$ because both shuffling and $db2$ mitigate the load imbalance between different $k$ indices in GEMM operation. Therefore, when the shuffling is used in Sparse.B designs, the impact of $db2$ get diminished as shown in the speedup results from B(2,$db2$,0, on/off).
  \item Balancing $db2$ and $db3$ is more effective than using only one large parameter between them because the speedup gain from $db2$ and $db3$ gets saturated as they increases. For example, B(2,1,1,on):2.6$\times$ outperform B(2,2,0,on):\\2.4$\times$ and B(2,0,2,on):2.4$\times$.
  \item Shuffler can boost the performance of nonzero $db2$ with a lower cost as shown in Figure~\ref{fig:W_configs}(b).
\end{enumerate}

\begin{table}[t]
\centering
\vspace{0pt}
\caption{Optimal design points and their routing configurations.}
\vspace{-5pt}
\label{tab:optimal_designs}
\footnotesize
\renewcommand{\arraystretch}{1.1} 
\resizebox{\columnwidth}{!}{%

\begin{tabular}{|c|c||c|c|c||c|c|c||c|}
\hline
\multicolumn{2}{|c||}{\multirow{2}{*}{Architecture}} & \multicolumn{3}{c||}{A-matrix Routing} & \multicolumn{3}{c||}{B-matrix Routing} & \multirow{2}{*}{Shuffle} \\ \cline{3-8}
\multicolumn{2}{|c||}{}                              & $da1$       & $da2$      & $da3$      & $db1$       & $db2$      & $db3$      &                          \\ \hline\hline
\multicolumn{2}{|c||}{Sparse.B*}                     & \multicolumn{3}{c||}{-}                & 4           & 0          & 1          & \multirow{6}{*}{On}      \\ \cline{1-8}
\multicolumn{2}{|c||}{Sparse.A*}                     & 2 & 1          & 0          & \multicolumn{3}{c||}{-}                &                          \\ \cline{1-8}
\multicolumn{2}{|c||}{Sparse.AB}                     & 2           & 0          & 0          & 2           & 0          & 1          &                          \\ \cline{1-8}
\multirow{3}{*}{Griffin}          & conf.B          & \multicolumn{3}{c||}{-}                & 8           & 0          & 1          &                          \\ \cline{2-8}
                                  & conf.A          & 2           & 1          & 1          & \multicolumn{3}{c||}{-}                &                          \\ \cline{2-8}
                                  & conf.AB         & 2           & 0          & 0          & 2           & 0          & 1          &                          \\ \hline
\end{tabular}
}
\vspace{-15pt}
\end{table}

Figure~\ref{fig:W_configs}(b),(c) show the power and area efficiency of weight-only sparse architecture families, for both pruned models \ins{(shown in y-axes)} and non-pruned models \ins{(shown in x-axes)}.  Looking into the Pareto optimal design points, Sparse.B(4,0,0,on), Sparse.B(4,0,1,on), and Sparse.B(4,0,2,on) show 95\%, 127\%, and 130\%  increase in power efficiency for pruned network, respectively.
These three designs only impose 10\%, 16\%, and 22\% power overhead respectively, compared to dense baseline models. (Listed in Table~\ref{tab:optimal_designs}) Among the three designs, we chose Sparse.B(4,0,1,on) as an optimal design point for weight-sparse architecture, Sparse.B*, which shows high $TOPS/W$ on DNN.B with minimal efficiency loss in DNN.dense. 
Finally, note that while SparTen.B shows 3.9$\times$ speedup, it hurts the power efficiency by 26\% and increases only 1\% area efficiency with DNN.B, compared to the baseline, due to large buffers, MUXs, and control path logic.
On the other hand, TCL.B improves both area and power efficiency.
However, we found that adding shuffling and $db3>0$ can significantly increase power efficiency for TCL.B up to 47\%.
 
\begin{figure*}[!th]
\vspace{-10pt}
\begin{center}
   \includegraphics[width=15.5cm]{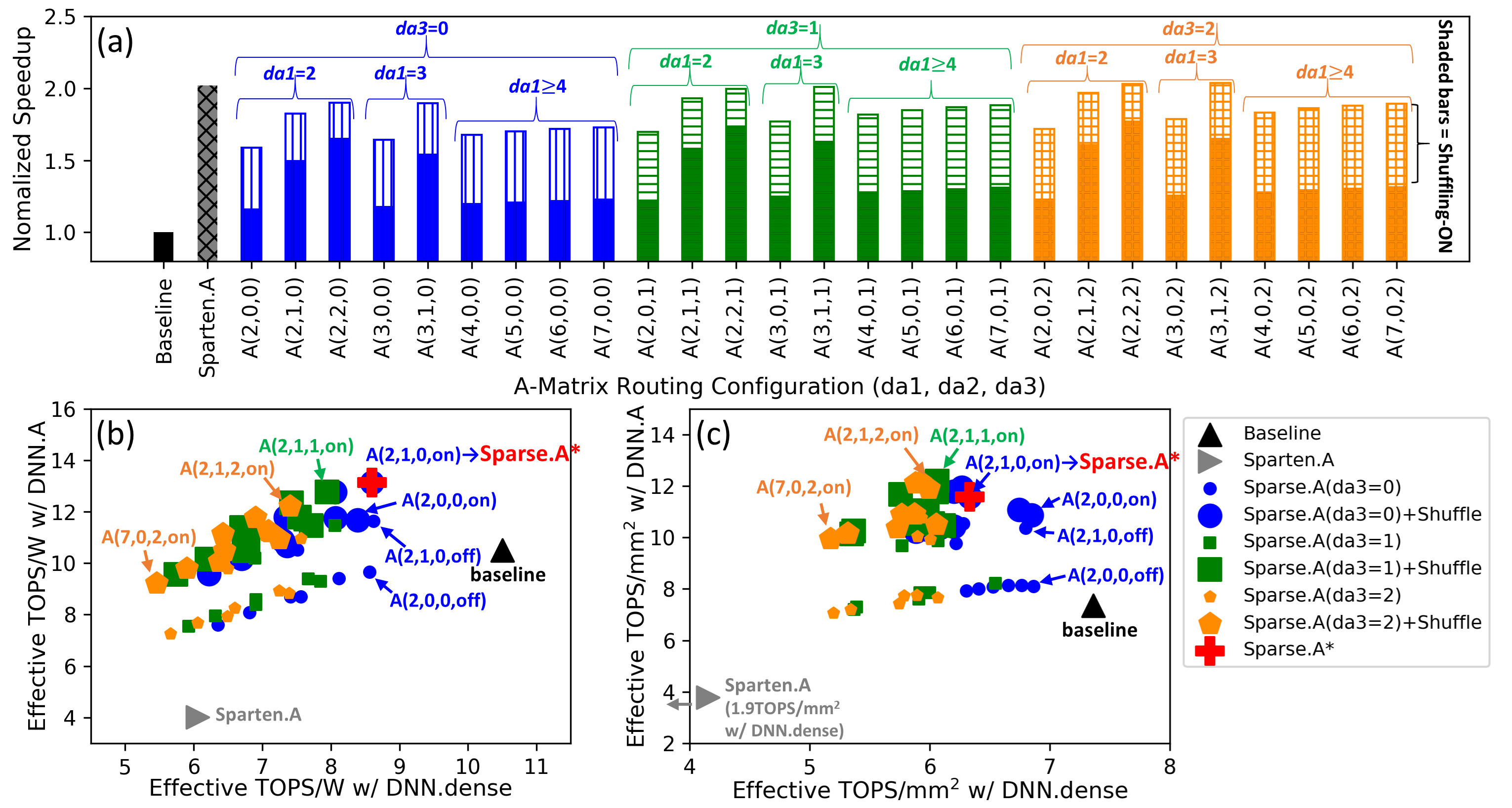}
\end{center}
\vspace{-15pt}
   \caption{
   Impact of A-matrix routing configurations. We use shorter notation A($da1$,$da2$,$da3$) to refer to Sparse.A($da1$,$da2$,$da3$). When we indicate shuffling-on/off of the sparse design, `on' or `off' index is added as Sparse.A($da1$,$da2$,$da3$,$on/off$). (a) Normalized speedup with respect to the dense baseline for different configurations. The shaded bars are the results with shuffling.
   (b) Effective power efficiency for different design points for DNN.A benchmark (y-axis) and DNN.dense benchmark (x-axis).
   (c) Effective area efficiency for different design points for DNN.A benchmark (y-axis) and DNN.dense benchmark (x-axis).  
}
\vspace{-15pt}
\label{fig:A_configs}
\end{figure*}

\subsection{Activation-Only Sparsity Support:}

We narrow down the design space exploration of architectures that support activation-only sparsity to those with AMUX/BMUX fan-in smaller than or equal to eight. This is based on our observation that designs with larger AMUX/\\BMUX also require deeper BBUF, which is expensive and reduces power and area efficiency.
Figure~\ref{fig:A_configs}(a) shows the designs with the above mentioned restriction. 
Figures~\ref{fig:A_configs}(b),(c) also show the power and area efficiency of these design points for a design with \ins{(shown in y-axes)} and without ReLU \ins{(shown in x-axes)}.
We observe that, 

\begin{enumerate}[leftmargin=*, label={(\arabic*)}]
  \item  $da1$ is not as important as $db1$ in Sparse.B architectures because the average sparsity level due to ReLU is close to 50\% that gives ideal speedup of $\sim$2x.(A(2,1,0,on):1.83$\times$\\$\sim$~A(3,1,0,on):1.89$\times$)
  \item $da3>0$ leads to power and area efficiency drop with insignificant speedup  (A(2,1,0,on):1.83$\times$ $\rightarrow$ A(2,1,1,on):\\1.93$\times$ $\rightarrow$ A(2,1,2,on):1.97$\times$) \ins{, while power/area overhead for A(2,1,1,on) and A(2,1,2,on) are 9\%/4\% and 17\%/6\%, respectively.}
  \item Shuffling boosts performance by up to 40\%(A(4,0,1,off):\\1.28$\times$ $\rightarrow$ A(4,0,1,on):1.79$\times$).
  \item Performance of designs with $da1\geq 4$ is limited since they cannot use $da2>0$ due to the AMUX fan-in size limit (\emph{cf.}AMUX=$1+da1\times(1+da2)\times(1+da3)$).
\end{enumerate}
Sparse.A(2,1,0,on), which we selected as an optimal design point (Sparse.A*) among activation sparse architectures, leads to 26\% increase in power efficiency for DNN.A models while reducing power efficiency by 18\% for DNN.dense models. (Listed in Table~\ref{tab:optimal_designs}) This configuration leads to 58\% increase in area efficiency, for DNN.A with only 14\% decrease for DNN.dense models.

We also observe that SparTen.A can achieve 2$\times$ speedup at the cost of 62\% and 49\% power and area overheads, compared to the optimized dense baseline. 
This is because SparTen.A does not unroll in $K$ dimensions, which leads to high accumulation cost and high operand fetch energy. 
In addition, only 8.5\% of the entire area of the SparTen.A is allocated to the compute units, which translates to low area efficiency (3.8 TOPS/mm$^2$).  

\begin{figure*}[ht!]
\vspace{-15pt}
\begin{center}
   \includegraphics[width=15.5cm]{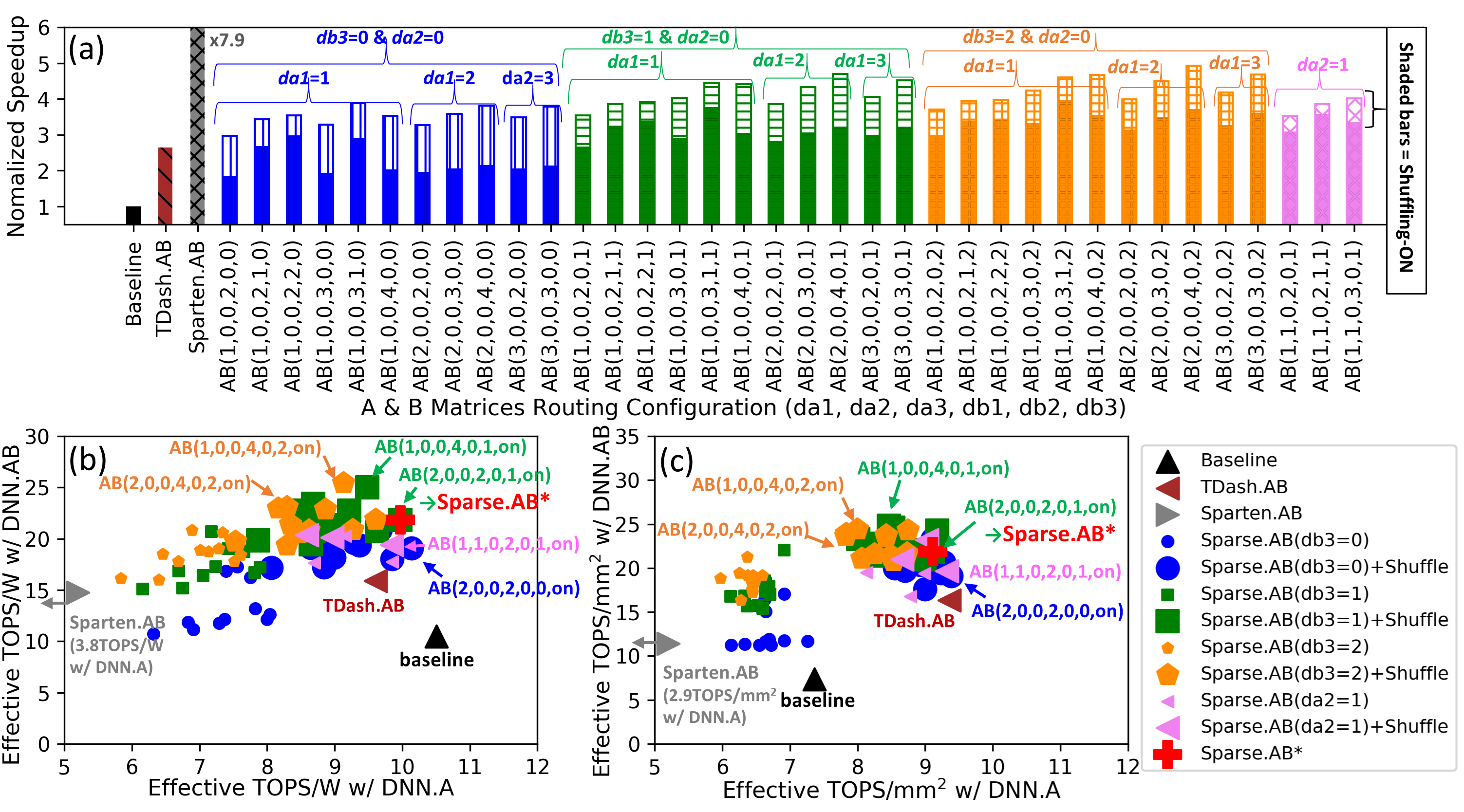}
\end{center}
\vspace{-15pt}
   \caption{
    Impact of A $\&$ B matrices routing configurations. We use shorter notation $AB(da1,da2,da3,db1,db2,db3)$ to refer to Sparse.AB $(da1,da2,da3,db1,db2,db3)$. When we indicate shuffling-on/off of the sparse design, `on' or `off' index is added as Sparse.AB$(da1,da2,da3,db1,db2,db3,on/off)$. (a) Normalized speedup with respect to the dense baseline for different configurations. The shaded bars are the results with shuffling.
    (b) Effective power efficiency for different design points for DNN.AB benchmark (y-axis) and DNN.A benchmark (x-axis), 
    (c) Effective Area efficiency for different design points for DNN.AB benchmark (y-axis) and DNN.A benchmark (x-axis).
}
\vspace{-10pt}
\label{fig:AW_configs}
\end{figure*}

\subsection{Dual Sparsity Support:}

The family of dual sparse architectures has seven parameters, namely $da1$, $da2$, $da3$, $db1$, $db2$, $db3$, and \emph{shuffling}. 
To explore the design space we allow configurations that lead to AMUX fan-in that is not larger than 16.
Compared to weight-only and activation-only sparsity, we consider larger fan-in, as dual sparse design can achieve higher performance, hence, can tolerate higher area and power overhead.
Several design points satisfy this restriction and we show the best-performing ones in Figure~\ref{fig:AW_configs}(a).
We observe that 4.9$\times$ speedup is achieved by Sparse.AB(2,0,0,4,0,2,on) compared to the dense baseline.
Figure~\ref{fig:AW_configs}(b)/(c) depict the projected power/area efficiency for DNN.A and DNN.AB models, on the X and Y axis, respectively.
We observe the following trends in Figure~\ref{fig:AW_configs}.
\begin{enumerate}[leftmargin=*, label={(\arabic*)}]
  \item  Shuffling can replace $db2>0$ and $da2>0$, which only imposes the sparsity overhead for DNN.dense by $\sim$2\% and $\sim$3\% for power and area, respectively. For example, Sparse.AB(1,$db2=0$,0,3,$db2=0$,1,on) shows 4.0$\times$ speedup while no-shuffling designs with $da2=1$ or $db2=1$ perform worse with 3.4$\times$ or 3.8$\times$ speedup, respectively.  
  \item If both $da3$ and $db3$ are nonzero, the design would not be on the Pareto optimal curve of power and area efficiency, as such designs require at least four adder trees per PE imposing a large overhead. 
  \item Most area and power-efficient designs have $da1\leq2$ as larger $da1$ results in larger BBUFs and AMUX/BMUX fan-in. Therefore, to achieve higher performance, it is effective to limit the maximum $da1$ window less than or equal to 2 and invest more on the weight side. In addition, $da3$ also increases AMUX fan-in size in contrast to $db3$. Thus, we chose $db3$ over $da3$ and excluded designs with $da3>0$ from Figure~\ref{fig:AW_configs} for sake of space.
\end{enumerate}

In all the explored design points, we selected Sparse.AB (2,0,0,2,0,1,on) as our optimal design point,  Sparse.AB*, as it shows 3.9$\times$ speedup and increases power and area efficiency of the dense baseline by 108\% and 187\% for DNN.AB models, respectively. (Listed in Table~\ref{tab:optimal_designs})
For DNN.A models, this design reduces power efficiency by only 5\% and improves area efficiency by 23\%. On the other hand, TDash.AB and Sparten.AB only improve power efficiency by 43\% and 40\%, respectively.
Both architectures do not exploit the benefits of weight preprocessing which can save the BBUF depth, BMUX fan-in size, and control overheads.

\subsection{Hybrid Architecture (Griffin):}

\begin{figure*}[t]
\vspace{-10pt}
\begin{center}
   \includegraphics[width=\linewidth]{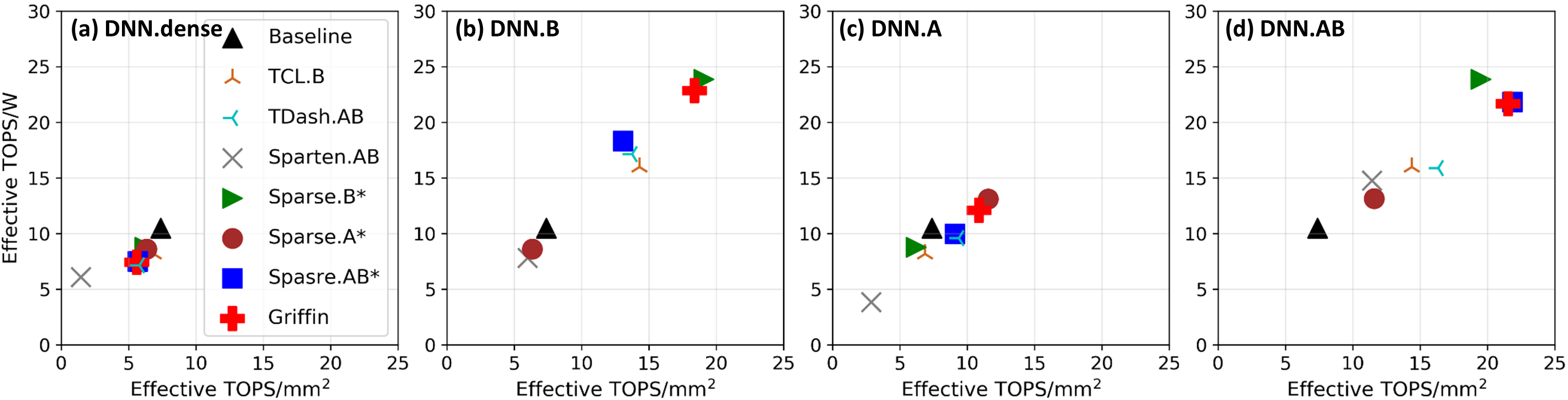}
\end{center} 
\vspace{-10pt}
   \caption{
   Power efficiency vs Area efficiency in different DNN categories such as (a) DNN.dense, (b)DNN.B, (c)DNN.A, (d)DNN.AB. 
}
\vspace{-10pt}
\label{fig:overall}
\end{figure*}

To evaluate our hybrid architecture, Griffin, we consider the impact of hybrid enhancement on Sparse.AB*. 
When running DNN.B models, it is reconfigured to  \emph{conf.B}$(db1,db2,db3,$ $shuffle)$ $=(8,0,1,on)$ with 3.5$\times$ speedup. 
This configuration of Griffin allows DNN.B models to achieve 25\% and 42\% better power efficiency and area efficiency, respectively, compared to Sparse.AB*, as shown in Figure~\ref{fig:overall}(b). 
The architecture uses \emph{conf.A}$(da1,da2,da3, shuffle)=(2,1,1,on)$, for running DNN.A models, with 1.94$\times$ speedup, which translates to 23\% and 20\% power efficiency and area efficiency improvement, respectively (Figure~\ref{fig:overall}(c)). The efficiency improvement in DNN.B and DNN.A was achieved by only $\sim$1\% drop of power efficiency and area efficiency in both DNN.dense and DNN.AB models ( Figure~\ref{fig:overall}(a) and (d)). All three configurations of Griffin are listed in Table~\ref{tab:optimal_designs}.
Griffin is also 1.2, 3.0, 3.1, and 1.4$\times$ more power-efficient and 3.8, 3.1, 3.7, and 1.8$\times$ more area-efficient compared to Sparten, the state-of-the-art dual sparse architecture to the best of our knowledge, for DNN.dense, DNN.B, DNN.A, and DNN.AB, respectively.
Note that the \emph{conf.A} of Griffin for DNN.A is not as effective as \emph{conf.B} for DNN.B, since activation tensors are denser than weight tensors.
Furthermore, activation sparsity is handled in real-time which is less effective than the preprocessing used for weight-only sparsity cases.

\subsection{Hardware Overhead and Breakdown:}
Table~\ref{tab:breakdown} shows the power and area breakdown of the dense baseline as well as Sparse.B*, TCL.B, Sparse.A*, Sparse.AB*, Griffin, TDash.AB, and Sparten.AB in the order of increasing power efficiency.
In the dense baseline, we observe that multipliers are dominant both in power and area. 
Allocating most of the resources to the compute unit makes the design more efficient and sparsity overhead more amplified.
In both Sparse.B* and Sparse.A* sparsity cases, the control overhead is insignificant. 
\ins{Moreover, power and area overhead of shuffler is less than 1.0\% and 0.7\% compared to the dense, respectively.}
In all types of sparse designs, the main power overheads come from registers added to the data path pipeline and those needed for ABUFs and BBUFs. 
The selected design for Sparse.A*, Sparse.B*, Sparse.AB*, and Griffin increase the data path power, compared to the dense baseline, by 46\%, 34\%, 72\%, and 73\%, respectively.
Moreover, these designs increase the area overhead with respect to the dense baseline, by 16\%, 19\%, 30\%, and 32\%, respectively.
We also observe that the control path imposes 12\% and 4.3\% power and area overhead, in both Sparse.AB* and Griffin, as they require one control unit per PE.

Since TCL.B and TDash.AB added sparse logic designed on top of efficient 3D dense core like Griffin, the power and area cost of those architectures are similar to Sparse.B* and Griffin, respectively. Therefore, the higher speedup from Sparse.B* and Griffin caused better power and area efficiency compared to TCL.B and TDash.AB. In the case of Sparsten.AB, both the PE and BUFs cause inefficiency because the Sparten.AB with MAC-based architecture does not share accumulators (which consume 110mW) and uses BUFs with a depth of 128. This buffer makes the power and area increase by 416mW and $6.4\times 10^5\mu m^2$, respectively (larger than the power and area of baseline architecture).

\begin{table*}[]
\centering
\caption{Power and area breakdown. CTRL and SHF are control-unit and shuffler, respectively, that are separated from PE. The results from PE are composed of REG/WR(pipeline registers and wires), ACC(accumulators), MUL(multipliers), ADT(adder trees), and MUXs. ASRAM and BSRAM results are merged into SRAM columns.}
\vspace{-5pt}
\label{tab:breakdown}
\setlength{\tabcolsep}{2pt} 
\renewcommand{\arraystretch}{1.5} 
\resizebox{0.95\textwidth}{!}{%

\begin{tabular}{|c||c|c|c|c|c|c|c|c|c|c|c||c|c|c|c|c|c|c|c|c|c|c|}

\hline
\multirow{3}{*}{Architecture}        & \multicolumn{11}{c||}{Power (mW)}                                                                                                                             & \multicolumn{11}{c|}{Area   ($\times$1000$\mu$m$^2$)}                                                                                                         \\ \cline{2-23} 
                                     & \multirow{2}{*}{Total} & \multirow{2}{*}{CTRL} & \multirow{2}{*}{SHF} & \multicolumn{2}{c|}{BUF} & \multicolumn{5}{c|}{PE}           & \multirow{2}{*}{SRAM} & \multirow{2}{*}{Total} & \multirow{2}{*}{CTRL} & \multirow{2}{*}{SHF} & \multicolumn{2}{c|}{BUF} & \multicolumn{5}{c|}{PE}            & \multirow{2}{*}{SRAM} \\ \cline{5-11} \cline{16-22}
                                     &                        &                       &                      & ABUF        & BBUF       & REG/WR  & ACC  & MUL  & ADT  & MUX   &                       &                        &                       &                      & ABUF         & BBUF      & REG/WR & ACC  & MUL & ADT  & MUX   &                       \\ \hline
Baseline                             & 151                    & -                     & -                    & -           & -          & 22.8 & 10.9 & 62.6 & 21.8 & -     & 33.3                  & 217                    & -                     & -                    & -            & -         & 3.2    & 2.6  & 29  & 6.7  & -     & 176                   \\ \hline
\textbf{Sparse.B*}  & 206                    & -                     & 0.7                  & 7.5         & -          & 41.0 & 10.9 & 55.4 & 20.4 & 3.5   & 66.7                  & 258                    & -                     & 0.9                  & 2.0          & -         & 4.0    & 2.6  & 33  & 12.8 & 6.5   & 196                   \\ \hline
TCL.B                                & 209                    & -                     & -                    & 4.3         & -          & 24.3 & 10.9 & 85.9 & 21.2 & 4.8   & 57.2                  & 233                    & -                     & -                    & 0.9          & -         & 3.4    & 2.6  & 34  & 6.6  & 6.3   & 179                   \\ \hline
\textbf{Sparse.A*}  & 223                    & 1.2                   & 0.4                  & 4.5         & 17.8       & 23.2 & 10.9 & 67.2 & 17.8 & 1.5   & 78.2                  & 253                    & 0.7                   & 0.5                  & 0.9          & 3.8       & 3.8    & 2.6  & 34  & 6.6  & 3.5   & 196                   \\ \hline
\textbf{Sparse.AB*} & 282                    & 18.2                  & 1.4                  & 15.3        & 22.9       & 64.5 & 10.9 & 31.7 & 17.8 & 7.0   & 92.3                  & 282                    & 8.1                   & 1.6                  & 11.5       & 5.2       & 6.0    & 2.6  & 29  & 12.3 & 17.5  & 188                   \\ \hline
\textbf{Griffin}    & 284                    & 18.2                  & 1.4                  & 15.3        & 22.9       & 64.5 & 10.9 & 31.7 & 17.8 & 8.8   & 92.3                  & 286                    & 9.4                   & 1.6                  & 11.5      & 5.2       & 6.0    & 2.6  & 29  & 12.3 & 20.7  & 188                   \\ \hline
TDash.AB                             & 284                    & 19.0                  & -                    & 5.8         & 23.4       & 24.3 & 10.9 & 85.9 & 21.2 & 9.6   & 84.1                  & 276                    & 8.9                   & -                    & 1.4          & 5.8       & 3.4    & 2.6  & 34  & 6.6  & 17.4  & 196                   \\ \hline
Sparten.AB                           & 991                    & 133                   & -                    & 213         & 213        & 7.5  & 110  & 133  & -    & inBUF & 181.6                 & 1139                   & 227                   & -                    & 320          & 320       & 0.7    & 30.2 & 41  & -    & inBUF & 200                   \\ \hline
\end{tabular}

}
\vspace{-15pt}
\end{table*}

\subsection{Overall Comparison:}
Figure~\ref{fig:overall} depicts the normalized area and energy efficiency of the several selected architectural configurations as well as the related work for four different types of DNN models. 
The goal for optimal design is to remain a top performer for all four categories of DNN models.
This is only achieved by Griffin design.
For DNN.dense models, most of the designs, except for the Sparten, perform in the same ballpark of efficiency and all less efficient of the dense baseline. 
That is expected as DNN.dense models do not reflect the gains in speedup associated with supporting sparsity. 
\ins{The reduced efficiency in Figure~\ref{fig:overall}(a) can be interpreted as \emph{sparsity tax} spent for the sake of the sparsity gain. Griffin shows more economic \emph{sparsity tax} of 29\%/24\% for power/area efficiency than 42\%/80\% of Sparten.AB.} 
However, these various designs show large contrast, especially in DNN.B and DNN.AB models
and beat the baseline by factors of integer in area/energy efficiency. 
The exceptions are Sparten.AB and Sparse.A* still perform overall worse than the baseline if the task is not dual sparse. 
TCL.B is close to the baseline for DNN.dense and DNN.A models and performs reasonably well for DNN.B and DNN.AB cases. 
While TDash.AB shows similar efficiency to Sparse.AW* in DNN.dense, DNN.A, and DNN.B, it does not perform well in 
DNN.AB models, compared to Sparse.AB* and Griffin.
\section{Related Work}
\label{sec:Related}

Recently, several optimization techniques are proposed to improve compute and memory efficiency for DNNs~\cite{deng2020model,liang2021pruning,sun2020sparse}. Various deep learning accelerators improve DNNs performance by exploiting weight and/or activation sparsity~\cite{dave2020hardware}.
Cambricon-X~\cite{CambriconX} and Bit-Tactical (TCL)~\cite{BitTactical} are two designs based on inner product compute units that focus only on exploiting weight sparsity. 
Cambricon-X leverages a queue for weights generated by flattening ($K$) axes and it uses a wide activation crossbar to fetch inputs corresponding with nonzero weights.
This approach routes nonzero weights with a large $16\times16$ window (i.e., $db1 = 16, db2=16$) which imposes high overhead in input crossbar and bandwidth.
This activation bandwidth requirement makes the design infeasible for scaling up. 
TCL introduces static scheduling coupled with a lightweight input multiplexing network.
TCL compresses weights statically by routing nonzero weights in time and input channel  (i.e., $db1$ and $db2$). 
TCL is similar to our weight-only sparsity design, but it does not support routing nonzero weights from different output channels (i.e.,$db3 = 0$).
Unlike weight sparsity, activation sparsity inherently exists in DNNs because of rectified linear (ReLU) function. 
Cnvlutin exploits only activation sparsity, without shuffling, by compressing them in time (i.e., $db1$)~\cite{Cnvlutin}. 

Other work proposed CNN architectures to support dual sparsity~\cite{SCNN, Sparten, Zena}. 
SCNN uses a 2D array of compute units that perform an outer product between two nonzero weight and activation vectors~\cite{SCNN}.  The outputs of the compute units are routed to their corresponding accumulator using a heavyweight crossbar which introduces a substantial overhead.   

ZeNA~\cite{Zena} is a dual sparsity architecture similar to Sparten. 
Both are based on sparse-aware MAC units which route computations in time (i.e., $db1$ and $da1$) only for each MAC unit independently. 
However, they are limited in efficiency and scalability due to their limited data movement and reuse between MACs.
In contrast, Griffin routes irregular sparsity among all core dimensions.

Eyeriss~v2~\cite{Eyerissv2} introduces a hierarchical mesh network for flexible processing of sparse weights and input activations directly in a compressed domain. In contrast to our work, it introduces large overheads degrading the performances of dense models. 
Sparse-TPU~\cite{sparse-TPU} incorporates index matching and value holding functionalities to efficiently process sparse matrices in a compressed format using a systolic array. 

In addition, there are other sparse architectures which exploit bit-level sparsity in weights and/or activation such as Stripe~\cite{bit-serial}, Bit-Pragmatic~\cite{bit-pragmatic}, Loom~\cite{loom}, Laconic~\cite{laconic}, Bit-Fusion~\cite{bit-fusion}, UNUP~\cite{unup} and activation-side in TCL~\cite{BitTactical}. 
Bit-level sparsity requires extra overhead due to bit-serial operation which degrades the power/area efficiency compared to bit-parallel designs. 

 Outerspace~\cite{outerspace} proposes a design based on cores to perform sparse outer products.
 SpArch~\cite{SpArch} optimizes sparse GEMM by reducing memory footprint (DRAM access).
Tensaurus~\cite{Tensaurus} introduces a new sparse storage format, compressed interleaved sparse slice (CISS) for sparse-dense matrix operations. 
ExTensor~\cite{ExTensor} targets general sparse tensor algebra, optimizing the memory hierarchy at multiple levels through a general abstraction based on intersections between non-zero data coordinates. 
Spaghetti~\cite{SPAGHETTI} designs a pattern-aware software scheduler to leverage sparsity patterns for optimal DRAM utilization.
Sparse CNN FPGA accelerator~\cite{structured_sparse} proposes a sparse dataflow to skip zero weights and minimize off-chip memory accesses. 
Procrustes~\cite{Procrustes} produces pruned models from the sparse DNNs training with an optimized Dropback algorithm. 
SIGMA~\cite{SIGMA} introduces a highly Flexible Dot Product Engine and Forward Adder Network to enable efficient GEMM computations in DNNs training.

\section{Conclusions}
\label{sec:conclusion}

This work describes a systematic approach to model the family of architectures that support various flavors of DNNs sparsity.
We explore the design space and offer multiple insights regarding the best dimensions to borrow from for varieties of sparse architectures.
We propose a hybrid architecture, Griffin, to enhance dual sparse architectures to perform close to optimal power efficiency even when both weight and activation tensors are not sparse.
We evaluate the proposed design in contrast with previous work for different model categories (i.e., DNN.dense, DNN.B, DNN.A, and DNN.AB). The result shows that Griffin architecture improves power and area efficiency by up to 3.1 and 3.8$\times$, respectively, compared to the state-of-the-art dual sparse architectures.


\clearpage


\bibliographystyle{IEEEtranS}
\bibliography{refs}

\begin{thebibliography}{10}
\providecommand{\url}[1]{#1}
\csname url@samestyle\endcsname
\providecommand{\newblock}{\relax}
\providecommand{\bibinfo}[2]{#2}
\providecommand{\BIBentrySTDinterwordspacing}{\spaceskip=0pt\relax}
\providecommand{\BIBentryALTinterwordstretchfactor}{4}
\providecommand{\BIBentryALTinterwordspacing}{\spaceskip=\fontdimen2\font plus
\BIBentryALTinterwordstretchfactor\fontdimen3\font minus
  \fontdimen4\font\relax}
\providecommand{\BIBforeignlanguage}[2]{{%
\expandafter\ifx\csname l@#1\endcsname\relax
\typeout{** WARNING: IEEEtranS.bst: No hyphenation pattern has been}%
\typeout{** loaded for the language `#1'. Using the pattern for}%
\typeout{** the default language instead.}%
\else
\language=\csname l@#1\endcsname
\fi
#2}}
\providecommand{\BIBdecl}{\relax}
\BIBdecl

\bibitem{armEthosN77}
\BIBentryALTinterwordspacing
Arm {E}thos-{N}77. [Online]. Available:
  \url{https://www.arm.com/products/silicon-ip-cpu/ethos/ethos-n77}
\BIBentrySTDinterwordspacing

\bibitem{nvdla}
\BIBentryALTinterwordspacing
{NVIDIA} deep learning accelerator {(NVDLA)}. [Online]. Available:
  \url{http://nvdla.org/primer.html}
\BIBentrySTDinterwordspacing

\bibitem{SPAGHETTI}
\BIBentryALTinterwordspacing
{SPAGHETTI}. [Online]. Available: \url{https://github.com/sfu-arch/SpGEMM}
\BIBentrySTDinterwordspacing

\bibitem{DC}
\BIBentryALTinterwordspacing
Synopsys design compiler. [Online]. Available:
  \url{https://www.synopsys.com/implementation-and-signoff/rtl-synthesis-test/design-compiler-graphical.html}
\BIBentrySTDinterwordspacing

\bibitem{v100}
\BIBentryALTinterwordspacing
(2017) {NVIDIA} {T}esla {V}100 {GPU} {A}chitecture. [Online]. Available:
  \url{https://images.nvidia.com/content/volta-architecture/pdf/voltaarchitecture-whitepaper.pdf}
\BIBentrySTDinterwordspacing

\bibitem{bit-pragmatic}
\BIBentryALTinterwordspacing
J.~Albericio, A.~Delm\'{a}s, P.~Judd, S.~Sharify, G.~O'Leary, R.~Genov, and
  A.~Moshovos, ``Bit-pragmatic deep neural network computing,'' in
  \emph{Proceedings of the 50th Annual IEEE/ACM International Symposium on
  Microarchitecture}, ser. MICRO-50 '17.\hskip 1em plus 0.5em minus 0.4em\relax
  New York, NY, USA: Association for Computing Machinery, 2017, p. 382–394.
  [Online]. Available: \url{https://doi.org/10.1145/3123939.3123982}
\BIBentrySTDinterwordspacing

\bibitem{Cnvlutin}
J.~Albericio, P.~Judd, T.~Hetherington, T.~Aamodt, N.~E. Jerger, and
  A.~Moshovos, ``Cnvlutin: Ineffectual-neuron-free deep neural network
  computing,'' \emph{ACM SIGARCH Computer Architecture News}, vol.~44, no.~3,
  pp. 1--13, 2016.

\bibitem{diannao}
\BIBentryALTinterwordspacing
T.~Chen, Z.~Du, N.~Sun, J.~Wang, C.~Wu, Y.~Chen, and O.~Temam, ``Diannao: A
  small-footprint high-throughput accelerator for ubiquitous
  machine-learning,'' in \emph{Proceedings of the 19th International Conference
  on Architectural Support for Programming Languages and Operating Systems},
  ser. ASPLOS '14.\hskip 1em plus 0.5em minus 0.4em\relax New York, NY, USA:
  Association for Computing Machinery, 2014, p. 269–284. [Online]. Available:
  \url{https://doi.org/10.1145/2541940.2541967}
\BIBentrySTDinterwordspacing

\bibitem{eyeriss}
\BIBentryALTinterwordspacing
Y.-H. Chen, J.~Emer, and V.~Sze, ``Eyeriss: A spatial architecture for
  energy-efficient dataflow for convolutional neural networks,'' in
  \emph{Proceedings of the 43rd International Symposium on Computer
  Architecture}, ser. ISCA '16.\hskip 1em plus 0.5em minus 0.4em\relax IEEE
  Press, 2016, p. 367–379. [Online]. Available:
  \url{https://doi.org/10.1109/ISCA.2016.40}
\BIBentrySTDinterwordspacing

\bibitem{Eyerissv2}
Y.-H. Chen, T.-J. Yang, J.~Emer, and V.~Sze, ``Eyeriss v2: A flexible
  accelerator for emerging deep neural networks on mobile devices,'' \emph{IEEE
  Journal on Emerging and Selected Topics in Circuits and Systems}, vol.~9,
  no.~2, pp. 292--308, 2019.

\bibitem{DaDianNao}
Y.~Chen, T.~Luo, S.~Liu, S.~Zhang, L.~He, J.~Wang, L.~Li, T.~Chen, Z.~Xu,
  N.~Sun \emph{et~al.}, ``Dadiannao: A machine-learning supercomputer,'' in
  \emph{2014 47th Annual IEEE/ACM International Symposium on
  Microarchitecture}.\hskip 1em plus 0.5em minus 0.4em\relax IEEE, 2014, pp.
  609--622.

\bibitem{dave2020hardware}
S.~Dave, R.~Baghdadi, T.~Nowatzki, S.~Avancha, A.~Shrivastava, and B.~Li,
  ``Hardware acceleration of sparse and irregular tensor computations of ml
  models: A survey and insights,'' \emph{arXiv preprint arXiv:2007.00864},
  2020.

\bibitem{BitTactical}
A.~Delmas~Lascorz, P.~Judd, D.~M. Stuart, Z.~Poulos, M.~Mahmoud, S.~Sharify,
  M.~Nikolic, K.~Siu, and A.~Moshovos, ``Bit-tactical: A software/hardware
  approach to exploiting value and bit sparsity in neural networks,'' in
  \emph{Proceedings of the Twenty-Fourth International Conference on
  Architectural Support for Programming Languages and Operating Systems}, 2019,
  pp. 749--763.

\bibitem{deng2020model}
L.~Deng, G.~Li, S.~Han, L.~Shi, and Y.~Xie, ``Model compression and hardware
  acceleration for neural networks: A comprehensive survey,'' \emph{Proceedings
  of the IEEE}, vol. 108, no.~4, pp. 485--532, 2020.

\bibitem{devlin2019bert}
J.~Devlin, M.-W. Chang, K.~Lee, and K.~Toutanova, ``Bert: Pre-training of deep
  bidirectional transformers for language understanding,'' 2019.

\bibitem{rigl}
\BIBentryALTinterwordspacing
U.~Evci, T.~Gale, J.~Menick, P.~S. Castro, and E.~Elsen, ``Rigging the lottery:
  Making all tickets winners,'' in \emph{Proceedings of the 37th International
  Conference on Machine Learning}, ser. Proceedings of Machine Learning
  Research, H.~D. III and A.~Singh, Eds., vol. 119.\hskip 1em plus 0.5em minus
  0.4em\relax PMLR, 13--18 Jul 2020, pp. 2943--2952. [Online]. Available:
  \url{http://proceedings.mlr.press/v119/evci20a.html}
\BIBentrySTDinterwordspacing

\bibitem{gale2019prune}
T.~Gale, E.~Elsen, and S.~Hooker, ``The state of sparsity in deep neural
  networks,'' \emph{arXiv preprint arXiv:1902.09574}, 2019.

\bibitem{Sparten}
A.~Gondimalla, N.~Chesnut, M.~Thottethodi, and T.~Vijaykumar, ``Sparten: A
  sparse tensor accelerator for convolutional neural networks,'' in
  \emph{Proceedings of the 52nd Annual IEEE/ACM International Symposium on
  Microarchitecture}, 2019, pp. 151--165.

\bibitem{caffe_con_troll}
S.~Hadjis, F.~Abuzaid, C.~Zhang, and C.~Ré, ``Caffe con troll: Shallow ideas
  to speed up deep learning,'' 2015.

\bibitem{DeepCompression}
S.~Han, H.~Mao, and W.~J. Dally, ``Deep compression: Compressing deep neural
  networks with pruning, trained quantization and huffman coding,'' \emph{arXiv
  preprint arXiv:1510.00149}, 2015.

\bibitem{han2015learning}
S.~Han, J.~Pool, J.~Tran, and W.~J. Dally, ``Learning both weights and
  connections for efficient neural networks,'' in \emph{Proceedings of the 28th
  International Conference on Neural Information Processing Systems - Volume
  1}, ser. NIPS'15.\hskip 1em plus 0.5em minus 0.4em\relax Cambridge, MA, USA:
  MIT Press, 2015, p. 1135–1143.

\bibitem{HassibiBabak1992Pruning}
B.~Hassibi and D.~G. Stork, ``Second order derivatives for network pruning:
  Optimal brain surgeon,'' in \emph{Proceedings of the 5th International
  Conference on Neural Information Processing Systems}, ser. NIPS'92.\hskip 1em
  plus 0.5em minus 0.4em\relax San Francisco, CA, USA: Morgan Kaufmann
  Publishers Inc., 1992, p. 164–171.

\bibitem{He_2015_ICCV}
K.~He, X.~Zhang, S.~Ren, and J.~Sun, ``Delving deep into rectifiers: Surpassing
  human-level performance on imagenet classification,'' in \emph{Proceedings of
  the IEEE International Conference on Computer Vision (ICCV)}, December 2015.

\bibitem{resnet}
K.~He, X.~Zhang, S.~Ren, and J.~Sun, ``Deep residual learning for image
  recognition,'' in \emph{Proceedings of the IEEE Conference on Computer Vision
  and Pattern Recognition (CVPR)}, June 2016.

\bibitem{sparse-TPU}
\BIBentryALTinterwordspacing
X.~He, S.~Pal, A.~Amarnath, S.~Feng, D.-H. Park, A.~Rovinski, H.~Ye, Y.~Chen,
  R.~Dreslinski, and T.~Mudge, ``Sparse-tpu: Adapting systolic arrays for
  sparse matrices,'' in \emph{Proceedings of the 34th ACM International
  Conference on Supercomputing}, ser. ICS '20.\hskip 1em plus 0.5em minus
  0.4em\relax New York, NY, USA: Association for Computing Machinery, 2020.
  [Online]. Available: \url{https://doi.org/10.1145/3392717.3392751}
\BIBentrySTDinterwordspacing

\bibitem{ucnn}
K.~{Hegde}, J.~{Yu}, R.~{Agrawal}, M.~{Yan}, M.~{Pellauer}, and C.~{Fletcher},
  ``Ucnn: Exploiting computational reuse in deep neural networks via weight
  repetition,'' in \emph{2018 ACM/IEEE 45th Annual International Symposium on
  Computer Architecture (ISCA)}, 2018, pp. 674--687.

\bibitem{ExTensor}
\BIBentryALTinterwordspacing
K.~Hegde, H.~Asghari-Moghaddam, M.~Pellauer, N.~Crago, A.~Jaleel, E.~Solomonik,
  J.~Emer, and C.~W. Fletcher, ``Extensor: An accelerator for sparse tensor
  algebra,'' in \emph{Proceedings of the 52nd Annual IEEE/ACM International
  Symposium on Microarchitecture}, ser. MICRO '52.\hskip 1em plus 0.5em minus
  0.4em\relax New York, NY, USA: Association for Computing Machinery, 2019, p.
  319–333. [Online]. Available: \url{https://doi.org/10.1145/3352460.3358275}
\BIBentrySTDinterwordspacing

\bibitem{gelu}
D.~Hendrycks and K.~Gimpel, ``Gaussian error linear units (gelus),''
  \emph{arXiv preprint arXiv:1606.08415}, 2016.

\bibitem{crane}
J.-W. Jang, S.~Lee, D.~Kim, H.~Park, A.~S. Ardestani, Y.~Choi, C.~Kim, Y.~Kim,
  H.~Yu, H.~Abdel-Aziz \emph{et~al.}, ``Sparsity-aware and re-configurable npu
  architecture for samsung flagship mobile soc,'' in \emph{Proceedings of the
  48th Annual International Symposium on Computer Architecture (ISCA)}, June
  2021.

\bibitem{tpu}
\BIBentryALTinterwordspacing
N.~P. Jouppi, C.~Young, N.~Patil, D.~Patterson, G.~Agrawal, R.~Bajwa, S.~Bates,
  S.~Bhatia, N.~Boden, A.~Borchers, R.~Boyle, P.-l. Cantin, C.~Chao, C.~Clark,
  J.~Coriell, M.~Daley, M.~Dau, J.~Dean, B.~Gelb, T.~V. Ghaemmaghami,
  R.~Gottipati, W.~Gulland, R.~Hagmann, C.~R. Ho, D.~Hogberg, J.~Hu, R.~Hundt,
  D.~Hurt, J.~Ibarz, A.~Jaffey, A.~Jaworski, A.~Kaplan, H.~Khaitan,
  D.~Killebrew, A.~Koch, N.~Kumar, S.~Lacy, J.~Laudon, J.~Law, D.~Le, C.~Leary,
  Z.~Liu, K.~Lucke, A.~Lundin, G.~MacKean, A.~Maggiore, M.~Mahony, K.~Miller,
  R.~Nagarajan, R.~Narayanaswami, R.~Ni, K.~Nix, T.~Norrie, M.~Omernick,
  N.~Penukonda, A.~Phelps, J.~Ross, M.~Ross, A.~Salek, E.~Samadiani, C.~Severn,
  G.~Sizikov, M.~Snelham, J.~Souter, D.~Steinberg, A.~Swing, M.~Tan,
  G.~Thorson, B.~Tian, H.~Toma, E.~Tuttle, V.~Vasudevan, R.~Walter, W.~Wang,
  E.~Wilcox, and D.~H. Yoon, ``In-datacenter performance analysis of a tensor
  processing unit,'' \emph{SIGARCH Comput. Archit. News}, vol.~45, no.~2, p.
  1–12, Jun. 2017. [Online]. Available:
  \url{https://doi.org/10.1145/3140659.3080246}
\BIBentrySTDinterwordspacing

\bibitem{bit-serial}
P.~{Judd}, J.~{Albericio}, T.~{Hetherington}, T.~M. {Aamodt}, and
  A.~{Moshovos}, ``Stripes: Bit-serial deep neural network computing,'' in
  \emph{2016 49th Annual IEEE/ACM International Symposium on Microarchitecture
  (MICRO)}, 2016, pp. 1--12.

\bibitem{Zena}
D.~Kim, J.~Ahn, and S.~Yoo, ``Zena: Zero-aware neural network accelerator,''
  \emph{IEEE Design \& Test}, vol.~35, no.~1, pp. 39--46, 2017.

\bibitem{alexnet}
\BIBentryALTinterwordspacing
A.~Krizhevsky, I.~Sutskever, and G.~E. Hinton, ``Imagenet classification with
  deep convolutional neural networks,'' \emph{Commun. ACM}, vol.~60, no.~6, p.
  84–90, May 2017. [Online]. Available: \url{https://doi.org/10.1145/3065386}
\BIBentrySTDinterwordspacing

\bibitem{kwon2020maestro}
H.~{Kwon}, P.~{Chatarasi}, V.~{Sarkar}, T.~{Krishna}, M.~{Pellauer}, and
  A.~{Parashar}, ``Maestro: A data-centric approach to understand reuse,
  performance, and hardware cost of dnn mappings,'' \emph{IEEE Micro}, vol.~40,
  no.~3, pp. 20--29, 2020.

\bibitem{Yann1989BrainDamage}
Y.~Le~Cun, J.~S. Denker, and S.~A. Solla, ``Optimal brain damage,'' in
  \emph{Proceedings of the 2nd International Conference on Neural Information
  Processing Systems}, ser. NIPS'89.\hskip 1em plus 0.5em minus 0.4em\relax
  Cambridge, MA, USA: MIT Press, 1989, p. 598–605.

\bibitem{lenet}
\BIBentryALTinterwordspacing
Y.~LeCun, B.~Boser, J.~S. Denker, D.~Henderson, R.~E. Howard, W.~Hubbard, and
  L.~D. Jackel, ``{Backpropagation Applied to Handwritten Zip Code
  Recognition},'' \emph{Neural Computation}, vol.~1, no.~4, pp. 541--551, 12
  1989. [Online]. Available: \url{https://doi.org/10.1162/neco.1989.1.4.541}
\BIBentrySTDinterwordspacing

\bibitem{lecun2015deep}
Y.~LeCun, Y.~Bengio, and G.~Hinton, ``Deep learning,'' \emph{nature}, vol. 521,
  no. 7553, pp. 436--444, 2015.

\bibitem{unup}
J.~{Lee}, C.~{Kim}, S.~{Kang}, D.~{Shin}, S.~{Kim}, and H.~{Yoo}, ``Unpu: An
  energy-efficient deep neural network accelerator with fully variable weight
  bit precision,'' \emph{IEEE Journal of Solid-State Circuits}, vol.~54, no.~1,
  pp. 173--185, 2019.

\bibitem{liang2021pruning}
T.~Liang, J.~Glossner, L.~Wang, and S.~Shi, ``Pruning and quantization for deep
  neural network acceleration: A survey,'' 2021.

\bibitem{davinci}
H.~{Liao}, J.~{Tu}, J.~{Xia}, and X.~{Zhou}, ``Davinci: A scalable architecture
  for neural network computing,'' in \emph{2019 IEEE Hot Chips 31 Symposium
  (HCS)}, 2019, pp. 1--44.

\bibitem{ma2020mobile}
\BIBentryALTinterwordspacing
X.~Ma, F.-M. Guo, W.~Niu, X.~Lin, J.~Tang, K.~Ma, B.~Ren, and Y.~Wang, ``Pconv:
  The missing but desirable sparsity in dnn weight pruning for real-time
  execution on mobile devices,'' \emph{Proceedings of the AAAI Conference on
  Artificial Intelligence}, vol.~34, no.~04, pp. 5117--5124, Apr. 2020.
  [Online]. Available:
  \url{https://ojs.aaai.org/index.php/AAAI/article/view/5954}
\BIBentrySTDinterwordspacing

\bibitem{maas2013rectifier}
A.~L. Maas, A.~Y. Hannun, and A.~Y. Ng, ``Rectifier nonlinearities improve
  neural network acoustic models,'' in \emph{Proc. icml}, vol.~30, no.~1.\hskip
  1em plus 0.5em minus 0.4em\relax Citeseer, 2013, p.~3.

\bibitem{TensorDash}
M.~{Mahmoud}, I.~{Edo}, A.~H. {Zadeh}, O.~{Mohamed Awad}, G.~{Pekhimenko},
  J.~{Albericio}, and A.~{Moshovos}, ``Tensordash: Exploiting sparsity to
  accelerate deep neural network training,'' in \emph{2020 53rd Annual IEEE/ACM
  International Symposium on Microarchitecture (MICRO)}, 2020, pp. 781--795.

\bibitem{tensorcore}
S.~{Markidis}, S.~W.~D. {Chien}, E.~{Laure}, I.~B. {Peng}, and J.~S. {Vetter},
  ``{NVIDIA} tensor core programmability, performance precision,'' in
  \emph{2018 IEEE International Parallel and Distributed Processing Symposium
  Workshops (IPDPSW)}, 2018, pp. 522--531.

\bibitem{MISHKIN201711}
\BIBentryALTinterwordspacing
D.~Mishkin, N.~Sergievskiy, and J.~Matas, ``Systematic evaluation of
  convolution neural network advances on the imagenet,'' \emph{Computer Vision
  and Image Understanding}, vol. 161, pp. 11--19, 2017. [Online]. Available:
  \url{https://www.sciencedirect.com/science/article/pii/S1077314217300814}
\BIBentrySTDinterwordspacing

\bibitem{ineffectualsource}
M.~{Nikolić}, M.~{Mahmoud}, A.~{Moshovos}, Y.~{Zhao}, and R.~{Mullins},
  ``Characterizing sources of ineffectual computations in deep learning
  networks,'' in \emph{2019 IEEE International Symposium on Performance
  Analysis of Systems and Software (ISPASS)}, 2019, pp. 165--176.

\bibitem{niu201926ms}
W.~Niu, X.~Ma, Y.~Wang, and B.~Ren, ``26ms inference time for resnet-50:
  Towards real-time execution of all dnns on smartphone,'' \emph{arXiv preprint
  arXiv:1905.00571}, 2019.

\bibitem{outerspace}
S.~{Pal}, J.~{Beaumont}, D.~{Park}, A.~{Amarnath}, S.~{Feng}, C.~{Chakrabarti},
  H.~{Kim}, D.~{Blaauw}, T.~{Mudge}, and R.~{Dreslinski}, ``Outerspace: An
  outer product based sparse matrix multiplication accelerator,'' in \emph{2018
  IEEE International Symposium on High Performance Computer Architecture
  (HPCA)}, 2018, pp. 724--736.

\bibitem{timeloop}
A.~{Parashar}, P.~{Raina}, Y.~S. {Shao}, Y.~H. {Chen}, V.~A. {Ying},
  A.~{Mukkara}, R.~{Venkatesan}, B.~{Khailany}, S.~W. {Keckler}, and J.~{Emer},
  ``Timeloop: A systematic approach to dnn accelerator evaluation,'' in
  \emph{2019 IEEE International Symposium on Performance Analysis of Systems
  and Software (ISPASS)}, 2019, pp. 304--315.

\bibitem{SCNN}
A.~Parashar, M.~Rhu, A.~Mukkara, A.~Puglielli, R.~Venkatesan, B.~Khailany,
  J.~Emer, S.~W. Keckler, and W.~J. Dally, ``Scnn: An accelerator for
  compressed-sparse convolutional neural networks,'' \emph{ACM SIGARCH Computer
  Architecture News}, vol.~45, no.~2, pp. 27--40, 2017.

\bibitem{park2016faster}
J.~Park, S.~Li, W.~Wen, P.~T.~P. Tang, H.~Li, Y.~Chen, and P.~Dubey, ``Faster
  cnns with direct sparse convolutions and guided pruning,'' \emph{arXiv
  preprint arXiv:1608.01409}, 2016.

\bibitem{Dark_Mem}
A.~{Pedram}, S.~{Richardson}, M.~{Horowitz}, S.~{Galal}, and S.~{Kvatinsky},
  ``Dark memory and accelerator-rich system optimization in the dark silicon
  era,'' \emph{IEEE Design Test}, vol.~34, no.~2, pp. 39--50, 2017.

\bibitem{LAP}
A.~Pedram, R.~A. van~de Geijn, and A.~Gerstlauer, ``Codesign tradeoffs for
  high-performance, low-power linear algebra architectures,'' \emph{IEEE
  Transactions on Computers}, vol.~61, no.~12, pp. 1724--1736, 2012.

\bibitem{prunedReluBert}
\BIBentryALTinterwordspacing
G.~Prato, E.~Charlaix, and M.~Rezagholizadeh, ``Fully quantized transformer for
  improved translation,'' \emph{CoRR}, vol. abs/1910.10485, 2019. [Online].
  Available: \url{http://arxiv.org/abs/1910.10485}
\BIBentrySTDinterwordspacing

\bibitem{SIGMA}
E.~{Qin}, A.~{Samajdar}, H.~{Kwon}, V.~{Nadella}, S.~{Srinivasan}, D.~{Das},
  B.~{Kaul}, and T.~{Krishna}, ``Sigma: A sparse and irregular gemm accelerator
  with flexible interconnects for dnn training,'' in \emph{2020 IEEE
  International Symposium on High Performance Computer Architecture (HPCA)},
  2020, pp. 58--70.

\bibitem{ramachandran2017swish}
P.~Ramachandran, B.~Zoph, and Q.~V. Le, ``Swish: a self-gated activation
  function,'' \emph{arXiv preprint arXiv:1710.05941}, 2017.

\bibitem{movementPruneBert}
\BIBentryALTinterwordspacing
V.~Sanh, T.~Wolf, and A.~Rush, ``Movement pruning: Adaptive sparsity by
  fine-tuning,'' in \emph{Advances in Neural Information Processing Systems},
  H.~Larochelle, M.~Ranzato, R.~Hadsell, M.~F. Balcan, and H.~Lin, Eds.,
  vol.~33.\hskip 1em plus 0.5em minus 0.4em\relax Curran Associates, Inc.,
  2020, pp. 20\,378--20\,389. [Online]. Available:
  \url{https://proceedings.neurips.cc/paper/2020/file/eae15aabaa768ae4a5993a8a4f4fa6e4-Paper.pdf}
\BIBentrySTDinterwordspacing

\bibitem{laconic}
S.~{Sharify}, A.~D. {Lascorz}, M.~{Mahmoud}, M.~{Nikolic}, K.~{Siu}, D.~M.
  {Stuart}, Z.~{Poulos}, and A.~{Moshovos}, ``Laconic deep learning inference
  acceleration,'' in \emph{2019 ACM/IEEE 46th Annual International Symposium on
  Computer Architecture (ISCA)}, 2019, pp. 304--317.

\bibitem{loom}
S.~{Sharify}, A.~D. {Lascorz}, K.~{Siu}, P.~{Judd}, and A.~{Moshovos}, ``Loom:
  Exploiting weight and activation precisions to accelerate convolutional
  neural networks,'' in \emph{2018 55th ACM/ESDA/IEEE Design Automation
  Conference (DAC)}, 2018, pp. 1--6.

\bibitem{bit-fusion}
H.~{Sharma}, J.~{Park}, N.~{Suda}, L.~{Lai}, B.~{Chau}, V.~{Chandra}, and
  H.~{Esmaeilzadeh}, ``Bit fusion: Bit-level dynamically composable
  architecture for accelerating deep neural network,'' in \emph{2018 ACM/IEEE
  45th Annual International Symposium on Computer Architecture (ISCA)}, 2018,
  pp. 764--775.

\bibitem{siustuart2018memory}
K.~{Siu}, D.~M. {Stuart}, M.~{Mahmoud}, and A.~{Moshovos}, ``Memory
  requirements for convolutional neural network hardware accelerators,'' in
  \emph{2018 IEEE International Symposium on Workload Characterization
  (IISWC)}, 2018, pp. 111--121.

\bibitem{Tensaurus}
N.~{Srivastava}, H.~{Jin}, S.~{Smith}, H.~{Rong}, D.~{Albonesi}, and
  Z.~{Zhang}, ``Tensaurus: A versatile accelerator for mixed sparse-dense
  tensor computations,'' in \emph{2020 IEEE International Symposium on High
  Performance Computer Architecture (HPCA)}, 2020, pp. 689--702.

\bibitem{sun2020sparse}
F.~Sun, M.~Qin, T.~Zhang, L.~Liu, Y.-K. Chen, and Y.~Xie, ``Computation on
  sparse neural networks and its implications for future hardware: Invited,''
  ser. DAC '20.\hskip 1em plus 0.5em minus 0.4em\relax IEEE Press, 2020.

\bibitem{mobilebert}
\BIBentryALTinterwordspacing
Z.~Sun, H.~Yu, X.~Song, R.~Liu, Y.~Yang, and D.~Zhou, ``{M}obile{BERT}: a
  compact task-agnostic {BERT} for resource-limited devices,'' in
  \emph{Proceedings of the 58th Annual Meeting of the Association for
  Computational Linguistics}.\hskip 1em plus 0.5em minus 0.4em\relax Online:
  Association for Computational Linguistics, Jul. 2020, pp. 2158--2170.
  [Online]. Available: \url{https://www.aclweb.org/anthology/2020.acl-main.195}
\BIBentrySTDinterwordspacing

\bibitem{attention_is_all}
\BIBentryALTinterwordspacing
A.~Vaswani, N.~Shazeer, N.~Parmar, J.~Uszkoreit, L.~Jones, A.~N. Gomez,
  L.~Kaiser, and I.~Polosukhin, ``Attention is all you need,'' in \emph{NIPS},
  2017, pp. 6000--6010. [Online]. Available:
  \url{http://papers.nips.cc/paper/7181-attention-is-all-you-need}
\BIBentrySTDinterwordspacing

\bibitem{wu2019pruneingedge}
\BIBentryALTinterwordspacing
R.-T. Wu, A.~Singla, M.~R. Jahanshahi, E.~Bertino, B.~J. Ko, and D.~Verma,
  ``Pruning deep convolutional neural networks for efficient edge computing in
  condition assessment of infrastructures,'' \emph{Computer-Aided Civil and
  Infrastructure Engineering}, vol.~34, no.~9, pp. 774--789, 2019. [Online].
  Available: \url{https://onlinelibrary.wiley.com/doi/abs/10.1111/mice.12449}
\BIBentrySTDinterwordspacing

\bibitem{Procrustes}
D.~{Yang}, A.~{Ghasemazar}, X.~{Ren}, M.~{Golub}, G.~{Lemieux}, and M.~{Lis},
  ``Procrustes: a dataflow and accelerator for sparse deep neural network
  training,'' in \emph{2020 53rd Annual IEEE/ACM International Symposium on
  Microarchitecture (MICRO)}, 2020, pp. 711--724.

\bibitem{yang2020interstellar}
\BIBentryALTinterwordspacing
X.~Yang, M.~Gao, Q.~Liu, J.~Setter, J.~Pu, A.~Nayak, S.~Bell, K.~Cao, H.~Ha,
  P.~Raina, C.~Kozyrakis, and M.~Horowitz, ``Interstellar: Using halide's
  scheduling language to analyze dnn accelerators,'' in \emph{Proceedings of
  the Twenty-Fifth International Conference on Architectural Support for
  Programming Languages and Operating Systems}, ser. ASPLOS '20.\hskip 1em plus
  0.5em minus 0.4em\relax New York, NY, USA: Association for Computing
  Machinery, 2020, p. 369–383. [Online]. Available:
  \url{https://doi.org/10.1145/3373376.3378514}
\BIBentrySTDinterwordspacing

\bibitem{SystemArdavan}
X.~Yang, J.~Pu, B.~B. Rister, N.~Bhagdikar, S.~Richardson, S.~Kvatinsky,
  J.~Ragan-Kelley, A.~Pedram, and M.~Horowitz, ``A systematic approach to
  blocking convolutional neural networks,'' \emph{arXiv preprint
  arXiv:1606.04209}, 2016.

\bibitem{CambriconX}
S.~Zhang, Z.~Du, L.~Zhang, H.~Lan, S.~Liu, L.~Li, Q.~Guo, T.~Chen, and Y.~Chen,
  ``Cambricon-x: An accelerator for sparse neural networks,'' in \emph{2016
  49th Annual IEEE/ACM International Symposium on Microarchitecture
  (MICRO)}.\hskip 1em plus 0.5em minus 0.4em\relax IEEE, 2016, pp. 1--12.

\bibitem{SpArch}
Z.~{Zhang}, H.~{Wang}, S.~{Han}, and W.~J. {Dally}, ``Sparch: Efficient
  architecture for sparse matrix multiplication,'' in \emph{2020 IEEE
  International Symposium on High Performance Computer Architecture (HPCA)},
  2020, pp. 261--274.

\bibitem{structured_sparse}
C.~{Zhu}, K.~{Huang}, S.~{Yang}, Z.~{Zhu}, H.~{Zhang}, and H.~{Shen}, ``An
  efficient hardware accelerator for structured sparse convolutional neural
  networks on fpgas,'' \emph{IEEE Transactions on Very Large Scale Integration
  (VLSI) Systems}, vol.~28, no.~9, pp. 1953--1965, 2020.

\bibitem{zhu2017prune}
M.~Zhu and S.~Gupta, ``To prune, or not to prune: exploring the efficacy of
  pruning for model compression,'' \emph{arXiv preprint arXiv:1710.01878},
  2017.

\end{thebibliography}

\end{document}